\documentclass{article}
\usepackage{ijcai26}

\usepackage{times}
\usepackage{soul}
\usepackage{url}
\usepackage[hidelinks]{hyperref}
\usepackage[utf8]{inputenc}
\usepackage[small]{caption}
\usepackage{graphicx}
\usepackage{amsmath}
\usepackage{amsthm}
\usepackage{booktabs}
\usepackage{algorithm}
\usepackage{algorithmic}
\usepackage{enumitem} 
\usepackage[switch]{lineno}

\usepackage{array}
\usepackage{booktabs}
\usepackage{xcolor}
\usepackage{pifont}
\usepackage{tikz}
\usepackage[utf8]{inputenc}
\usepackage[T1]{fontenc}
\usepackage{xurl}
\usepackage{amsmath}
\usepackage{amssymb}
\usepackage{mathtools}
\usepackage{amsthm}
\usepackage{bm}
\usepackage{multirow}
\usepackage{diagbox}
\usepackage{multicol}
\usepackage{microtype}
\usepackage{graphicx}
\usepackage{subcaption}  % 关键：提供 subfigure 环境
\usepackage{booktabs} % for professional tables
\usepackage{xcolor}
\definecolor{deepgreen}{rgb}{0,0.5,0}
\definecolor{deepred}{rgb}{1, 0, 0}    % 深红色（下降箭头专用，新定义 ✅ 推荐）

\newcommand{\mname}{\texttt{EnzyPGM}}
\title{\mname: Pocket-conditioned Generative Model for Substrate-specific\\
Enzyme Design}

\author{
Zefeng Lin$^1$
\and
Zhihang Zhang$^4$\and
Weirong Zhu$^{1}$\and
Tongchang Han$^1$\\
Xianyong Fang$^3$\thanks{Corresponding author, Email: fangxianyong@ahu.edu.cn}
\and
Tianfan Fu$^2$\thanks{Corresponding author, Email: futianfan@nju.edu.cn}
\and
Xiaohua Xu$^1$\thanks{Corresponding author, Email: xiaohuaxu@ustc.edu.cn}
\\
\affiliations
$^1$School of Computer Science and Technology, University of Science and Technology of China, Hefei, Anhui, China\\
$^2$State Key Laboratory for Novel Software Technology at Nanjing University, School of Computer Science, Nanjing University, Nanjing, Jiangsu, China\\
$^3$School of Computer Science and Technology, Anhui University, Hefei, Anhui, China\\
$^4$School of Airspace Science and Engineering, Shandong University, Jinan, Shandong, China\\
\emails
\{zflin, zwr211355,tchan,xiaohuaxu\}@mail.ustc.edu.cn,
futianfan@nju.edu.cn,\\
fangxianyong@ahu.edu.cn,
202200800170@mail.sdu.edu.cn
}
\begin{document}

\maketitle

\begin{abstract}
Designing enzymes with substrate-binding pockets is a critical challenge in protein engineering, as catalytic activity depends on the precise interaction between pockets and substrates. Currently, generative models dominate functional protein design but cannot model pocket-substrate interactions, which limits the generation of enzymes with precise catalytic environments. To address this issue, we propose \textbf{\mname}, a unified framework that jointly generates enzymes and substrate-binding pockets conditioned on functional priors and substrates, with a particular focus on learning accurate pocket–substrate interactions. At its core, \mname\ includes two main modules: a Residue-atom Bi-scale Attention (RBA) that jointly models intra-residue dependencies and fine-grained interactions between pocket residues and substrate atoms, and a Residue Function Fusion (RFF) that incorporates enzyme function priors into residue representations. Also, we curate {EnzyPock}, an enzyme–pocket dataset comprising {83,062} enzyme–substrate pairs across {1,036} four-level enzyme families. 
Extensive experiments demonstrate that \mname\ achieves state-of-the-art performance on EnzyPock. Notably, \mname\ reduces the average binding energy of 0.47\,kcal/mol over EnzyGen, showing its superior performance on substrate-specific enzyme design. The code and dataset will be released later.
% maintain 74.87 in pLDDT 优越性  with binding pockets  and structure
% . Notably, \mname\ obtains 20\% improvement over EnzyGen in pLDDT and recovers 70\% amino acids. in average binding energy
% \textcolor{red}{tianfan: we cannot highlight ablation results, instead we need to highlight improvement over best baseline (e.g., 20\% improvement over xxx baseline) or absolute accuracy (e.g., 70\% AA recovery rate)} textbf
% A joint training objective encompassing sequence reconstruction and structural consistency further ensures the generation of functionally valid enzymes.

\end{abstract}

\section{Introduction}

\begin{figure}[t]
\centering
\includegraphics[width=0.85\linewidth]{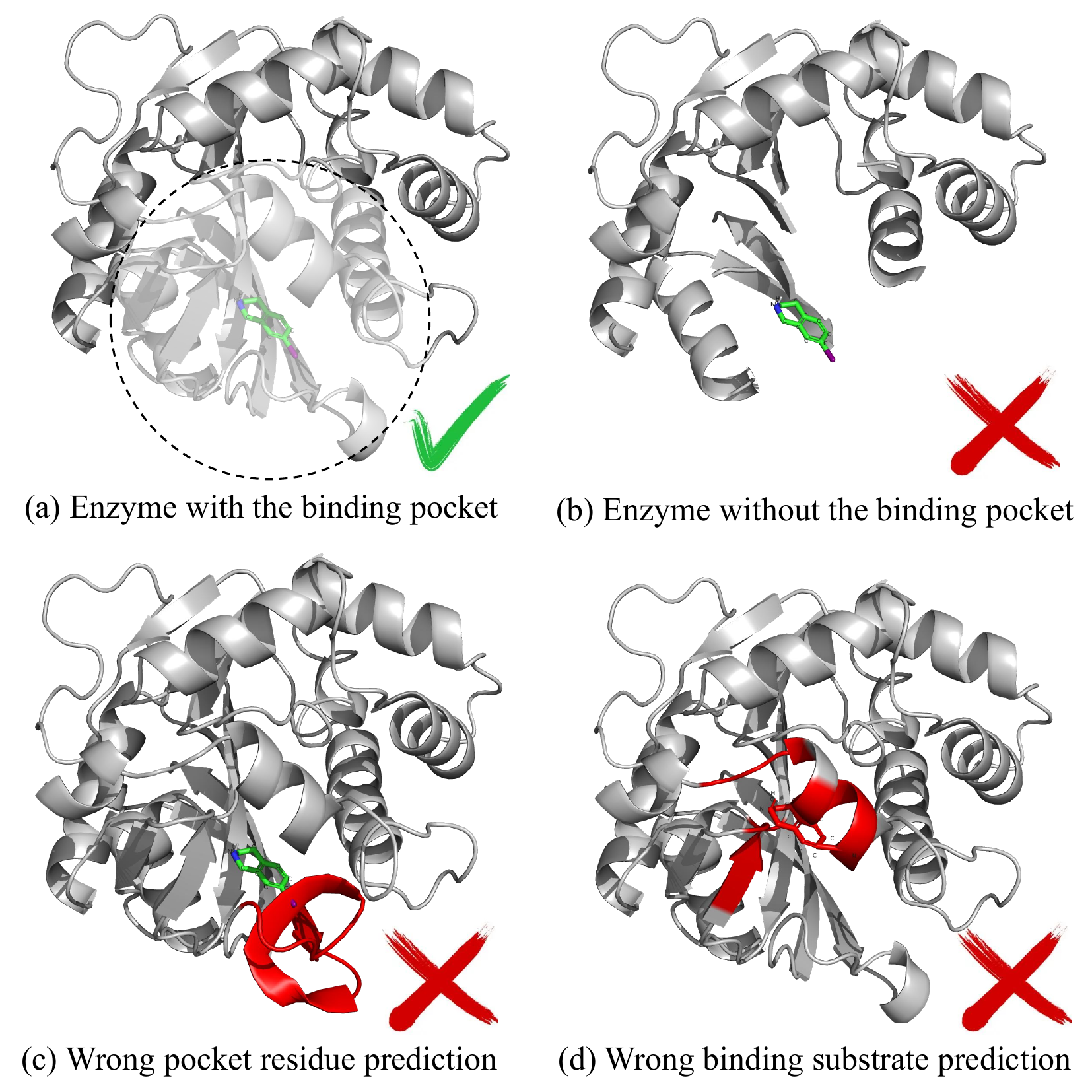}
\caption{
% 酶设计的四种情况，图(a)是真实带结合口袋的酶；图(b)是没有口袋的酶；图(c)是预测了错误口袋残基的酶；图(d)是预测了错误底物的酶。
Four cases of enzyme design. 
% (a): an enzyme with a real binding pocket; (b):  an enzyme without a pocket; (c): an enzyme with a predicted wrong pocket residue; (d): an enzyme with a predicted wrong substrate.  can 
None of the enzymes in (b), (c), and (d) bind to the substrate because there are no pockets in (b), and atom clashes in (c) and (d), which are represented by red.
}
\label{fig:contrast}
\end{figure}

Enzymes are nature’s most efficient biocatalysts, accelerating chemical reactions by $10^3$–$10^{17}$ fold \cite{Bar-Even2011} while maintaining unique substrate specificity~\cite{Huang2021}. 
The catalytic activity of enzymes arises from precisely folded 3D structures that form confined spatial regions, called substrate-binding pockets~\cite{stank2016protein}, where specific substrates are recognized, bound, and rapidly transformed into products. 
% Owing to their biological catalytic activity, enzymes play critical roles in various biological and industrial processes, including biosynthesis, pharmaceutical manufacturing, and sustainable chemical production. 
Recently, deep learning methods have demonstrated impressive potential for functional protein design~\cite{chuSparksFunctionNovo2024,kortemmeNovoProteinDesign2024,notinMachineLearningFunctional2024a}, where these methods can be classified as:  
% zhouNavigatingLandscapeEnzyme2024 chuSparksFunctionNovo2024,
(1) general functional protein design guided by functional labels and partial protein information~\cite{Anishchenko2021,Madani2022}; % Dauparas2022ProteinMPNN 
% wuProteinSequenceDesign2021 Tegge2022 
(2) ligand-binding protein design relies on the geometric structure of protein-ligand complexes~\cite{RFdiffusion2023,dauparasAtomicContextconditionedProtein2025a}; 
%  of enzyme-catalyzed reactions krishnaGeneralizedBiomolecularModeling2024
and (3) substrate-specific enzyme design based on the substrate properties and enzyme function~\cite{SENZ2025,wang2025artificial}. 
% ,songenzycontrol SENZ2025 hua2024enzymeflow

% 然而,联合建模酶和口袋仍然缺乏研究,因为它面临几个关键挑战 understudied
\noindent\textbf{Challenges.}
However, joint modeling of enzymes and substrate-binding pockets remains rarely studied as it faces several key challenges:
% However, jointly modeling substrate-binding pockets and enzymes faces several key challenges: 
(1) How to learn an effective representation of substrate-binding pockets to capture pocket-substrate interactions? These interactions affect the binding configuration and geometric structure of the enzyme and substrate~\cite{Stark2022EquiBind}.
% 图(b)中的案例展示了没有建模口袋-底物相互作用的情况，这时预测出来的酶没有结合口袋，因此无法与底物正常结合。 Yang2020 ,Gainza2020MaSIF
Figure~\ref{fig:contrast}(b) shows the case without the pocket-substrate interaction modeling, where the predicted enzyme does not have a binding pocket and cannot bind to the substrate.
(2) How to accurately model the residues and substrate in the pocket? The minor prediction deviations on residue and substrate significantly reduce catalytic efficiency~\cite{Lesk2010}.
% 图(c)中的案例展示了没有的情况，这时预测出来的酶没有结合口袋，因此无法与底物正常结合。 Anishchenko2021 Jumper2021AlphaFold2
% \ref{fig:contrast}  substrate-binding  in the pocket accurate
Figure \ref{fig:contrast}(c) and (d) show the cases without accurately modeling the residue and substrate, where the predicted pocket and substrate have an atom clash.
(3) A lack of public datasets that contain enzyme and substrate binding pockets. 
% The case in Figure \ref{fig:contrast}(d) shows the case without the accurate residue modeling in the pocket, where the predicted enzyme has a residue clash with the substrate. between
% 
% While the research community has released large enzyme databases~\cite{such as Brenda} and protein binding datasets~\cite{such as PDB Bind}, there is still a lack of enzyme-pocket refined datasets specifically designed for substrate binding pockets modeling.
% While the research community has made available comprehensive enzyme databases~\cite{Chang2021} and protein-ligand binding datasets~\cite{Wang2005}, there remains a lack of refined enzyme-pocket datasets specifically designed for modeling substrate binding pockets. structural geometric equivariance and fuse the functional priors 
% Enzyme Pocket-conditioned Generative Model

\noindent\textbf{Solution.}
To tackle these challenges, in this paper, we propose the \textbf{Pocket-conditioned Enzyme Generative Model (\mname)}, which explicitly models the substrate-binding pocket to design substrate-specific enzymes.
% 我们的关键设计动机在于, Enzyme
% the interaction of enzymes and substrates, as well as update representations and coordinates of  enzyme and 功能分类 类别 category develop a pocket-enhanced module that models pocket-substrate interactions, guiding guide the enzyme generative model based on the pocket-substrate interactions
Our key motivation is to model pocket-substrate interactions by bi-scale representation learning, thereby dynamically adjusting pocket sites in the enzyme to form a valid pocket.
Furthermore, given that enzymes have obvious functional categories and functionally conserved sites, we integrate them into the language model for enzyme design.
% 描述要具体一些 via masked language modeling are mainly determined by
% Specifically, for capturing interaction between the pocket residues and substrate atoms and accurately modeling residue and substrate atoms representations, we propose the Residue-atom Bilevel Attention (RBA) module. based on such biological priors
Specifically, we propose the Residue-atom Bi-scale Attention (RBA) module to accurately model the interaction between pocket residues and substrate atoms.
% a cross-modal bilevel attention mechanism is proposed to enable multi-scale representation learning across pockets and substrates.    
% 分别地 within the enzyme within the pocket
It mainly consists of intra-residue and residue-atom attention mechanisms, which individually capture neighborhood correlations among residues, and the interaction between pocket residues and substrate atoms, thus facilitating bi-scale representation learning across pockets and substrates.
% The motivation of this design is to realize cross-modal message exchange in residue and atom levels, thus facilitating multi-scale representation learning across pockets and substrates.
To fuse the enzyme function and the structure of residues into the residue representation, we employ the Residue Function Fusion (RFF) module. 
It integrates Enzyme Commission (EC) number as the function embedding and updates residue representations and 3D coordinates based on their spatial neighborhoods. 
% For ensure the SE(3)-SNE to model the substrate atom.
% Then we
% which is the Transformer block with a function embedding and Spatial Neighborhood Equivariant (SNE) layer, where the former integrates the Enzyme Commission (EC) number, and the latter updates residue representations based on their local 3D neighborhoods.   
% the Residue Function Fusion Pock
% Furthermore, to solve the lack of public datasets in enzyme-pocket modeling, we construct EnzyPock, an enzyme-pocket refined dataset of \textbf{84,336} enzyme-substrate pairs derived from PDBbind~\cite{wangPDBbindDatabaseCollection2004} and CrossDocked~\cite{francoeurThreeDimensionalConvolutionalNeural2020}. 
% It includes enzyme-substrate pairs across \textbf{17,404} PDB entries and \textbf{1,036} four-level enzyme families, which significantly fills the gap of joint modeling of enzymes and binding pockets in current research.
Also, to solve the lack of public datasets in enzyme-pocket modeling, we curate {EnzyPock}, the first enzyme–pocket dataset comprising {84,336} enzyme–substrate pairs across {1,036} four-level enzyme families and {17,404} PDB entries.
% 这样对吗
% 四级EC家族 train:16713 val:382 test:350
% During training, \mname\ optimizes a joint objective that integrates amino acid type prediction and structural reconstruction for enzyme and pocket residues, enabling the model to learn the sequence and structure of the binding pocket and enzymes.

\noindent\textbf{Main contributions} are summarized as:
% (1)
% (2)
(1) We propose \mname, the first model to generate enzyme and substrate-binding pockets conditioned on specific substrates jointly.
% , which takes the proposed ....  by two main modules, RBA and RFF. 
% (2) We design a residue-atom bi-scale attention mechanism for multi-scale representation learning, which captures the interaction across residues and substrate atoms to generate an accurate binding pocket.
% (2) We design the RBA module for bi-scale representation learning, which captures the interaction across residues and substrate atoms to generate an accurate binding pocket.
% novel insight 我们在融合了功能先验的酶生成模型基础上
% We are based on an enzyme generation model that integrates functional priors and adopt a function and fuse enzyme functional priors to design enzymes 并结合酶功能先验和底物设计 enzymes with an
(2) We design a bi-scale attention mechanism and fuse enzyme function priors, which accurately captures the interaction across residues and substrate atoms to generate a binding pocket.
(3) We curate EnzyPock, a refined enzyme-pocket dataset to tackle the lack of datasets for enzyme-pocket jointly modeling. 
Experimental results show that \mname\ consistently outperforms existing methods across on EnzyPock, generating enzymes with higher catalytic activity, structural confidence, and substrate selectivity.

\section{Related Work}
\label{sec:related}
\noindent\textbf{Deep Learning for Protein Design}. 
Recent years have witnessed significant progress in protein design driven by deep learning.
These protein design methods can be categorized into three paradigms: structure-based inverse folding, \textit{de novo} sequence generation, and joint sequence-structure design.
Inverse folding (IF) models aim to obtain amino acid sequences that fold into a given 3D structure. Representative methods, such as ProteinMPNN~\cite{Dauparas2022ProteinMPNN} and ESM-IF~\cite{Lin2023ESMIF}, achieve high natural sequence recovery but rely on a given protein backbone.
\textit{De novo} sequence generation methods aim at designing novel sequences from scratch.
They generate valid sequences by learning evolutionary patterns from large datasets, e.g., ProtGPT2~\cite{Ferruz2022ProtGPT2}, DPLM~\cite{Wang2024DPLM}, and DiMA~\cite{Wang2024DiMA}.
% ProGen2~\cite{Madani2022},
Sequence-structure co-design methods generate both sequence and structure jointly based on multi-modal priors, such as equivariant graph neural network (EGNN)~\cite{Satorras2021EGNN} and ESM3~\cite{Lin2024}. 
Although these approaches demonstrate strong potential for protein engineering, they fail to consider catalytic functions and substrate-specific requirements, which are critical to enzyme design.
% General functional protein design often fails to model enzyme-specific function and enzyme-substrate interactions accurately.   without structural priors structurally  emphasize intrinsic protein structural constraints and

\noindent\textbf{Ligand-binding Pocket Generation}. 
% To extend protein generation beyond intrinsic structure learning,   由于配体结合口袋在药物设计方面的重要性  to model protein–ligand interactions
Owing to the importance of the ligand-pocket binding in drug design, several studies propose ligand-aware pocket generation frameworks.
% For example, ligandMPNN \cite{dauparasAtomicContextconditionedProtein2025a} conditions sequence generation on ligand geometry, while RFdiffusionAA \cite{krishnaGeneralizedBiomolecularModeling2024} performs ligand-guided diffusion to generate protein structure binding to the given ligand.  Ligand-conditioned and further 
PocketGen \cite{zhangEfficientGenerationProtein2024a} and PocketFlow \cite{Li2024PocketGen_Generalized} explore protein pocket generation by integrating molecular priors to generate ligand-binding pockets.  
Despite these advances, most pocket generation models rarely consider the functional prior of proteins and require protein backbones as the generation condition. 
This limits these approaches to accurately capture protein–substrate interactions and apply to substrate-specific enzyme design. 
% Ligand-specific protein design rarely considers the functional prior of enzymes.
% % target the substrate-binding pocket of enzymes.  ligand-conditioned and 
% % explicitly considers the substrate and enzyme catalytic function, but enzyme-substrate specificity and 
% Substrate-specific enzyme design neglects the modeling of the substrate-binding pocket. 

\noindent\textbf{Substrate-specific Enzyme Generation}. 
EnzyGen \cite{EnzyGen2024} formulated enzyme generation as the joint modeling of enzyme sequences and backbones, conditioned on functional residues and substrates.
% , showing that substrate-aware conditioning improves the catalytic performance of generated enzymes.
% SENZ \cite{SENZ2025} further introduced zero-shot substrate-specified enzyme generation to synthesize enzymes capable of catalyzing reactions unseen during training.  分阶段设计口袋和酶
GENzyme~\cite{hua2024reaction} is a \textit{de novo} design approach that generates pockets, enzymes and substrate-binding complexes in stages.
EnzyControl~\cite{songenzycontrol} focuses on enzyme backbone generation by functional and substrate-specific control.
However, they do not explicitly model the pocket–substrate interactions, which makes them unable to generate enzymes with high substrate affinity.
% In contrast, the proposed \mname\ integrates substrate-binding pocket generation directly into the enzyme design, thereby enabling the design of practical substrate-specific enzymes. substrate-binding pocket and 
% In contrast, the proposed \mname\ jointly model enzyme and substrate-binding pocket, thereby generating practical substrate-specific enzymes.
In contrast, the proposed \mname\ integrates substrate-binding pocket modeling directly into the enzyme design, thereby generating practical substrate-specific enzymes.

% 先介绍酶和底物里面有哪些东西，其中序列的残基类型由20个氨基酸类型组成
% RBA模块的具体组成 these inputs
% the bi-scale attention  in RBA
% (a) \mname\ takes enzyme function-conversed sites and their coordinates, EC number, as well as substrate features and coordinates as inputs, processes them via RFF and RBA modules, and outputs the complete enzyme sequence, structure, and substrate-binding pocket. substrate-binding  as the condition 恢复的掩码位点用输出端用红色表示 
\begin{figure*}[t]
\centering
\includegraphics[width=0.95\linewidth]{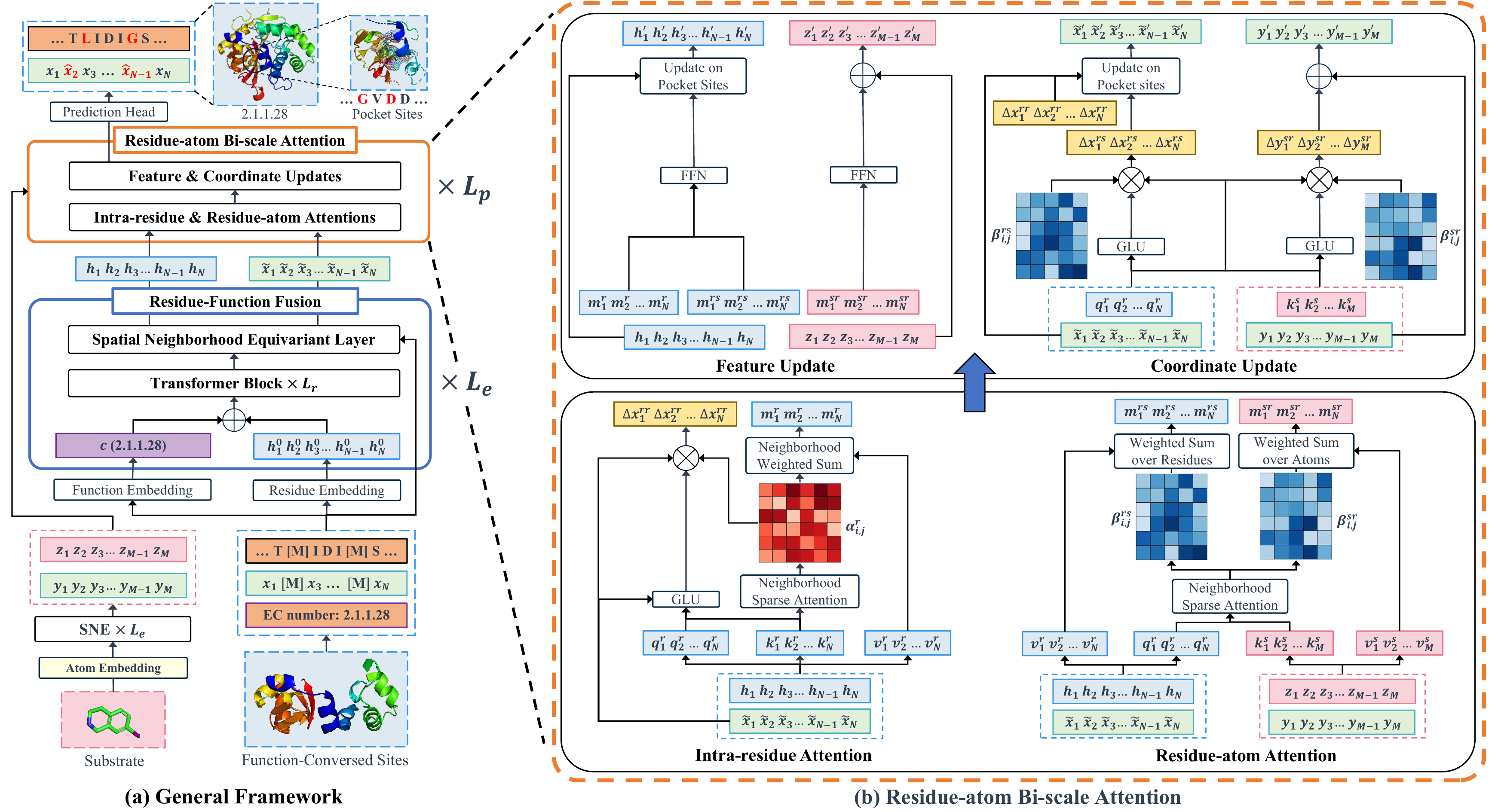} 
\caption{Overview of \mname. 
% (a) \mname\ takes enzyme function-conversed sites and their coordinates as a masked input sequence, conditioned on EC number and substrate features and coordinates, and predicts the complete enzyme sequence, structure, and pocket sites via RFF and RBA modules. 
(a) \mname\ takes enzyme function-conversed sites as a masked input sequence, conditioned on EC number and substrate atom, and predicts the complete enzyme and pocket sites via RFF and RBA modules.
The mask sites at the input are represented by [M] both in sequence and coordinate space, their prediction results are indicated in \textcolor{red}{\textbf{red}} at the output.
(b) present the specific process of the intra-residue and residue-atom attention, the feature and coordinate updates.
}
\label{fig:modeloverview}
\end{figure*}

\section{\mname: Pocket-conditioned Enzyme Generative Model}

\subsection{Substrate-specific Enzyme Design}

% 与之前的底物特定酶设计方法不同，我们的模型旨在同时建模酶和催化口袋，从而在生成过程中引入与底物结合的催化口袋的引导，以生成能够与底物更加紧密结合的酶。 Problem: 
% 以生成
% We formally define the task of \textit{substrate-specified enzyme design} as follows.   
% Unlike previous substrate-specific enzyme design approaches~\cite{EnzyGen2024,SENZ2025}, our model 

% We aim to guide by bi-scale representation learning
% This work aims at modeling the substrate-binding pocket and the enzyme jointly, thereby generating enzymes with substrate-binding capability. 
This work aims to design substrate-specific enzymes with the binding pocket by modeling pocket-substrate interactions and fusing enzyme function priors.
% thereby obtaining enzymes that bind more tightly to the substrate. the residue coordinate  of $r_i$

An $N$-residue enzyme $\mathcal{E}$ is represented as a sequence of residues $\mathcal{E} = \{r_1, r_2, \dots, r_N\}$, where each residue $r_i = (s_i, x_i)$, with $s_i \in \mathcal{A}$ denoting the amino acid type, and $x_i \in \mathbb{R}^3$ being its 3D coordinate of the $C_\alpha$ atom. Here, $\mathcal{A}$ represents the 20-amino-acid alphabet.  
A substrate molecule $\mathcal{V}$ consisting of $M$ atoms is denoted as $\mathcal{V} = \{v_1, v_2, \dots, v_M\}$, where each atom $v_j = (a_j, y_j)$, with $a_j \in \mathbb{R}^{5}$ representing the five atom-level chemical feature composed of element type, aromaticity, connectivity, hydrogen atom number, and hybridization type, and $y_j \in \mathbb{R}^3$ being its 3D coordinate.
% (e.g., atom type, hybridization, aromaticity, formal charge) \textit{substrate-binding pocket}
% 5元组  \in \mathbb{R}^{5}  5-tuple enzyme  defined as 

Given an enzyme $\mathcal{E}$ and substrate $\mathcal{V}$, the substrate-binding pocket $\mathcal{P}$ is the residue set spatially proximal to the substrate:
\begin{equation}
    \mathcal{P} = \{r_i \in \mathcal{E} \mid \min_{v_j \in \mathcal{V}} \|x_i - y_j\|_2 < d\},
    \label{eq:pocketdef}
\end{equation}
where $\| \cdot \|_2$ denotes the L2 norm, and $d$ is a distance threshold (in \AA) and set to 10 in our practice.

% 这里讲述如何用MSA获取保守位点和EC标签C，要讲述获取保守位点的动机、逻辑和构造方式
% belonging to EC label $c$ in the EC tree 
% 为什么要获取EC标签c和保守位点： tipton2000history Webb1992 and classifying families 
% 为了充分利用上述生物学先验，指导生成更具靶向活性和实践应用价值的酶，我们参考已有研究方法 Enzyme Commission (EC) Bickel2002
Enzyme Commission (EC) numbers are a standard for enzymes based on catalytic reaction types, where the fourth-level classification precisely defines functional families and provides a basis for inferring enzyme functions via homology~\cite{tipton2000history}.
Functionally conserved sites of enzymes are highly conserved across evolution, as these residues directly determine catalytic activity and structural stability, serving as key signatures of enzymes' functions and structures~\cite{Tristem2000}.
% 
% 实用性 Practicality practical utility and utilize them to generate enzymes for generating enzymes with improved activity and practicality
To fully utilize these functional priors, we obtain conserved sites \(\mathcal{M}\) under EC label \(c\) as part of the input using multiple sequence alignment (MSA), and then predict other masked sites using a model. 
% Specifically, we 
% and mask other sites.
% and use them as the conditional masked inputs of the model. 
% 
% 按我们的流程进行MSA，收集保守位点数据
% 具体来说，我们为收集的酶-口袋数据集中的每条酶检索其在BRENDA中的EC编号或通过CLEAN模型预测其EC编号。然后我们将同一个4级EC编号对应的酶家族下的所有酶通过ClustalW2 tool进行MSA以获得该家族下的保守位点，其中Residues present at the same position in more than \(\tau\) sequences are defined as Functionally conserved sites。In practice, we set \(\tau\) to 30\% of the total sequences in the MSA.   from the BRENDA database~\cite{Schomburg2021} 
% 
% Specifically, we first retrieve the EC number for each enzyme in our collected enzyme-pocket dataset from the Uniprot\footnote{\url{https://www.uniprot.org/}} database.
% We then collect all enzymes corresponding to each fourth-level EC family and perform MSA on these enzymes using the ClustalW2 tool~\cite{Larkin2007} to identify conserved sites within the family. 
% In practice, we select residues present at the same position in more than 30\% sequences as the conserved sites.
% where the conserved sites are residues present at the same position in more than \(\tau\) sequences, as illustrated in Figure~\ref{appen:msa}.   
% set \(\tau\) to 30\% of the total sequences in the MSA.  

% The problem studied in this paper can be formulated as follows: given a desired enzyme family c in EC tree, a substrate V, a functionally important site index set M
% 功能保守位点 综上，我们的目标是 functionally and structurally consistent  a given substrate $\mathcal{V}$.
% Accordingly, of substrate-specific enzyme design the problem that we study  has high binding capability with
In summary, our study problem can be formulated as a conditional masked language modeling (MLM) task: given a EC label $c$, a given substrate $\mathcal{V}$, a functionally conserved site index set $\mathcal{M}$, its residues set $\mathcal{E}_\mathcal{M}$, generate a complete enzyme $\hat{\mathcal{E}}$ and its substrate-binding pocket $\hat{\mathcal{P}}$ that can bind the substrate.
Essentially, we need to train a generative model to learn the probability $P(\hat{\mathcal{E}}, \hat{\mathcal{P}} \mid \mathcal{E}_\mathcal{M}, \mathcal{V}, c)$.
% the goal of our model

\subsection{Overall Framework}

% 为了在给定ec分类和保守位点的条件下生成与给定底物结合的酶和口袋，我们提出了\mname，如图所示，它主要包括三个子层。其中，

% To generate enzyme and substrate-binding pockets conditioned on both enzyme functional class and functionally conserved sites, we propose the \textbf{Pocket-conditioned Enzyme Generative Model (\mname)}. As illustrated in  the goal on the \textbf{Spatial Neighborhood Equivariant (SNE)}, 
% Pocket-conditioned Enzyme Generative Model The \textbf{\mname}
% 前面加一段动机,动机按照重要性,结构按照input顺序展开
\mname\ is proposed for generating enzyme and substrate-binding pockets conditioned on both enzyme functional class and functionally conserved sites (Figure~\ref{fig:modeloverview} (a)). 
It mainly consists of two modules: the \textbf{Residue Function Fusion (RFF)} and the \textbf{Residue-atom Bi-scale Attention (RBA)}. 
% is a unified generative architecture designed to jointly model enzyme scaffolds, substrate-binding pockets, and substrate atom under geometric and functional constraints.  There are

% First, RFF embeds both residue and enzyme-class functional information into a unified representation space. and substrate molecular  by the function embedding and Transformer blocks  by the SNE layer then refines the residue representations and coordinates derived from
At a high level, RFF aims to fuse residue-level contextual information from the enzyme sequence with functional priors in EC number and maintain their SE(3)-equivariance, while RBA focuses on enhancing pocket residues in both feature and coordinate spaces selectively.
Next, we will detail the components of \mname.

\subsection{Residue–Function Fusion (RFF)}

RFF is stacked Transformer~\cite{Vaswani2017} blocks with a function embedding and Spatial Neighborhood Equivariant (SNE) layer. 
The former embeds the EC number into residue embeddings to obtain function-aware residue embeddings, while the latter performs message passing within the residue neighborhood graphs to ensure SE(3)-equivariance of residue representations and coordinates. 
We also use the SNE layer to model the substrate atom and coordinates.
% Specifically, it integrates sequence and EC number embedding to generate function-aware residue representations. enzyme family features

%  = \{(s_i, x_i)\}_{i=1}^N where $s_i$ denotes the amino acid type and $x_i \in \mathbb{R}^3$ the residue coordinate, , structure  + \text{Emb}(\text{$x_i$}) and $x_i$ an enzyme sequence  conserved site set
Given $\mathcal{E}$ and $\mathcal{M}$, each $s_i$ in $\mathcal{E}$ are first embedded into a residue  $\mathbf{h}_i^{(1)} \in \mathbb{R}^{d_h}$:
\begin{equation}
\begin{aligned}
    \mathbf{h}_i^{(1)} &=  \begin{cases}
    \text{Emb}(\text{$s_i$}), & i \in \mathcal{M}, 
    \\
    \text{Emb}(\text{[mask]}), & \text{otherwise}, \end{cases}
     \label{eq:rff_input_new}
\end{aligned}
\end{equation}
where [mask] is the special token to mask the sites we need to predict, and $d_h$ denotes the dimensions of the representation. 

% specific constraints hidden state
For introducing the enzyme family priors, we encode the EC number $c$ into a function embedding $\text{Emb}(c) \in \mathbb{R}^{d_h}$ using a learnable embedding table and add it to the residue embedding. 
% Then, we apply a global multi-head self-attention sub-layer (MHA) to model long-range residue dependencies and a fully connected feedforward network (FFN) to obtain the function-enhanced residue representation for the $t+1$-th transformer block ($t=0,1,\dots$): integrated
Then, we apply $L_r$ Transformer blocks with function embedding to model long-range residue dependencies. In the $t$-th transformer block, a multi-head self-attention (MHA) and fully connected feedforward network (FFN) are performed to obtain the function-aware residue representation $\mathbf{h}_i^{(t+1)}$:
% a global multi-head self-attention sub-layer (MHA) to model long-range residue dependencies and a fully connected feedforward network (FFN) to obtain the function-enhanced residue representation for the $t+1$-th transformer block ($t=0,1,\dots$):  
% 
% We then adopt the Transformer block~\cite{Vaswani2017} to obtain the function-enhanced residue representation.
% Specifically, we apply a global multi-head self-attention sub-layer (MHA) to model long-range residue dependencies and a fully connected feedforward network (FFN) to compute the representation. 
% For the $t+1$-th transformer block ($t=0,1,\dots$), the update of function-aware representations is expressed as:
% \begin{equation}
\begin{align}
    \tilde{\mathbf{h}}_i^{(t)} &= \mathbf{h}_i^{(t)} + \text{Emb}(c), \\
    \mathbf{h}_i^{(t+0.5)} &= \tilde{\mathbf{h}}_i^{(t)} + \text{MHA}\!\left(\text{LN}(\tilde{\mathbf{h}}_i^{(t)}), \text{LN}(\tilde{\mathbf{H}}^{(t)})\right), \\
    \mathbf{h}_i^{(t+1)} &= \mathbf{h}_i^{(t+0.5)} + \text{FFN}\!\left(\text{LN}(\mathbf{h}_i^{(t+0.5)})\right).
    \label{eq:rff_update_preln}
\end{align}
% \end{equation}
Here $\text{LN}(\cdot)$ denotes the layer normalization (LN) operator, Pre-LN instead of Post-LN~\cite{Xiong2020} is adopted, and $\tilde{\mathbf{H}}^{(t)} = [\tilde{\mathbf{h}}_1^{(t)}, \dots, \tilde{\mathbf{h}}_N^{(t)}]$. 
The output $\{\mathbf{h}_i^{(L_r+1)}\}_{i=1}^N$ of the last Transformer block and the residue coordinates are then input to the SNE layer to ensure their SE(3)-equivariance: 
% \begin{equation}
\begin{align}
    \mathbf{m}_{ik} &= \text{FFN}\!\left(\mathbf{h}_i^{(L_r+1)}; \mathbf{h}_k^{(L_r+1)}; \|x_i - x_k\|_2^2\right), \\
    \tilde{\mathbf{h}}_i &= \text{FFN}\!\left(\mathbf{h}_i^{(L_r+1)}; \sum_{k \in \mathcal{N}(i)} \mathbf{m}_{ik}\right), \\
    \tilde{x}_i &= x_i + \frac{1}{|\mathcal{N}(i)|} \sum_{k \in \mathcal{N}(i)} (x_i - x_k) \, \cdot \text{FFN}\!\left(\mathbf{m}_{ik}\right). 
    \label{eq:sne_coord_update}
\end{align}
% \end{equation}
Here, $\mathbf{m}_{ik}$ denotes messages exchanged between residues, $\mathcal{N}(i)$ is the local neighboring nodes set of residue $r_i$ within a predefined radius, $|\mathcal{N}(i)|$ represents the number of the set, and [;] denotes concatenation of vectors.  
% where $\mathcal{N}(i)$ and $\mathcal{N}(j)$ are the local neighborhoods of residue $r_i$ and atom $v_j$ within a predefined radius.
% Where $|\mathcal{N}(i)|$ represents the number of neighboring nodes within a predefined radius and is used for normalization.   $\tilde{\mathbf{h}}_i$ is the updated representation by SNE,
% 
% $\{\mathbf{z}_j, y_j\}$ and $\mathbf{u}_{jk}^{(l)}$  and atom pairs geometry-aware 
% In practice, we stack 1 layer of SNE on every 11 layers of RFF to ensure the SE(3)-equivariance.
Finally, the SNE layer outputs refined residue representations $\{\tilde{\mathbf{h}}_i\}_{i=1}^N$ with updated coordinates $\{\tilde{x}_i\}_{i=1}^N$, which are passed to the RBA module.
% that capture local spatial correlations while remaining spatial equivariance. and geometry-aware atom representations $\{\mathbf{z}_j\}_{j=1}^M$. for the pocket enhancement
% Then, the representations are passed to the next layer for the pocket enhancement.  

Similarly, we stack $L_e$ of SNE layers to learn the substrate atom representation $\{\mathbf{z}_j\}_{j=1}^M$ from the embedding of the $j$-th substrate atom chemical feature $\mathbf{z}_j^{(1)} = \text{Emb}(a_j)$, but we do not update the atom coordinates.
% as shown in Figure~\ref{fig:modeloverview} (a).
% \begin{equation}
% \begin{aligned}
%     % \mathbf{z}_j^{(1)} &= \text{Emb}(a_j)  \\
%      \mathbf{u}_{jk}^{(l)} &= \text{FFN}\!\left(\mathbf{z}_j^{(l)}; \mathbf{z}_k^{(l)}; \|y_j - y_k\|_2^2\right) \\
%     \mathbf{z}_j^{(l+1)} &= \text{FFN}\!\left(\mathbf{z}_j^{(l)}; \sum_{k \in \mathcal{N}(j)} \mathbf{u}_{jk}^{(l)}\right)  
%     \label{eq:sne_coord_update}
% \end{aligned}
% \end{equation}
% where $\mathbf{z}_j^{(1)} = \text{Emb}(a_j)$ is the embedding of the $j$-th substrate atom chemical feature and $\mathbf{z}_j^{(l+1)}$ is the representation at the $l$-th SNE layer. 
% Here, we do not update the atom coordinates and use the real ones for subsequent computation, while obtaining the substrate atom representation $\{\mathbf{z}_j\}_{j=1}^M$.
% After this, we obtained the substrate atom representation $\{\mathbf{z}_j\}_{j=1}^M$.
% with SE(3)-equivariance.
% following \cite{guan2023d}  In the SNE layer

\subsection{Residue-atom Bi-scale Attention (RBA)}  
% RBA layer aims to obtain residue representations by jointly modeling residue–residue and residue–substrate interactions, while selectively enhancing pocket residues in both feature and coordinate spaces. 
% After obtaining refined residue representations and coordinates, as well as substrate atom representations, we employ RBA to model pocket–substrate interactions to enhance the enzyme pocket.  as shown in 
% Specifically,  selectively
RBA aims to enhance pocket residues in both feature and coordinate spaces by jointly modeling (I) residue–residue and (II) residue–substrate interactions (Figure~\ref{fig:modeloverview}(b)). 
% to enhance the enzyme pocket features and coordinates. features and coordinates
% 
% Then, RBA updates the pocket residue representations and coordinates and substrate atom coordinates using the neighborhood messages of residues and the interactions between residues and atoms.
% % 
% Finally, RBA outputs a pocket-conditioned residue representation and the updated coordinates of the residue and substrate atom, thereby predicting enzyme and substrate-binding pockets.
% 
% RBA layer aims to selectively enhance pocket residues in both feature and coordinate spaces by jointly modeling (I) residue–residue and (II) residue–substrate interactions.
% 
% The layer primarily operates using a bilevel attention mechanism: intra-residue attention captures global message exchange within the enzyme, and residue–atom attention introduces pocket-substrate interaction. By integrating the bilevel attention, the pocket-conditioned residue representations and coordinates are finally obtained.
% 机制  (RBF)

\paragraph{(I) Intra-Residue Attention.}  
% Obtained the geometry-aware residue representations $\{\mathbf{h}_i\}_{i=1}^{N}$ and their coordinates $\{\tilde{x}_i\}_{i=1}^{N}$, the intra-residue attention integrates geometric proximity using a radial basis function $\phi_{\text{rbf}}$. Specifically, the attention weight between the $i$-th and $k$-th residues is computed as: geometry-aware $i$-th and $k$-th
% integrates geometric proximity 径向向量
To capture global message exchange within the enzyme, we perform the intra-residue attention to integrate intra-residue message using a radial basis function $\phi_{\text{rbf}}$ based on the refined residue representations $\{\tilde{\mathbf{h}}_i\}_{i=1}^{N}$ and coordinates $\{\tilde{x}_i\}_{i=1}^{N}$. 
Specifically, the attention weight $\alpha_{ik}^{r}$ between the residue $r_i$ and $r_k$ is calculated as:
\begin{equation}
\alpha_{ik}^{r} = \text{softmax}\!\left(  
\frac{(\mathbf{q}_i^{r})^\top \mathbf{k}_k^{r}}{\sqrt{d}} 
+ \mathbf{W}_{r}^\top \phi_{\text{rbf}}\!\left(\|\tilde{x}_i - \tilde{x}_k\|_2\right)
\right),
\end{equation}
where $\mathbf{q}_i^{r} = \mathbf{W}_Q^{r} \cdot \text{LN}(\tilde{\mathbf{h}}_i)$, $\mathbf{k}_k^{r} = \mathbf{W}_K^{r} \cdot \text{LN}(\tilde{\mathbf{h}}_k)$, and $\mathbf{v}_k^{r} = \mathbf{W}_V^{r} \cdot \text{LN}(\tilde{\mathbf{h}}_k)$ are the query, key, and value projections of residue representation, and $\mathbf{W}_{[\cdot]}$ is a linear projection weight.  
% where $\mathbf{q}_i^{r} = \mathbf{W}_Q^{r} \cdot \text{LN}\left(\tilde{\mathbf{h}}_i\right)$, $\mathbf{k}_k^{r} = \mathbf{W}_K^{r}\tilde{\mathbf{h}}_k$, and $\mathbf{v}_k^{r} = \mathbf{W}_V^{r}\tilde{\mathbf{h}}_k$ are the query, key, and value projections of residue features, respectively, and $\mathbf{W}_{[\cdot]}$ is a linear projection weight.  features, respectively the $i$-th attention-weighted

The intra-residue message $\mathbf{m}_i^{r}$ from neighboring residues and the coordinate update amount $\Delta x_i^{rr}$ for residue $r_i$ are then computed as:
% \begin{equation}
\begin{align}
    \mathbf{m}_i^{r} &= \sum_{k \in \mathcal{N}(i)} \alpha_{ik}^{r} \mathbf{v}_k^{r}, \\
    g_{ik}^r &= \text{GLU} \left( \mathbf{q}_i^r ; \mathbf{k}_k^r ; \phi_{\text{rbf}} \left(\|\tilde{x}_i - \tilde{x}_k\|_2\right) \right) \label{eq:gate}, \\
    \Delta x_i^{rr} &= \sum_{k \in \mathcal{N}(i)} \alpha_{ik}^{r} \cdot g_{ik}^r \cdot \frac{\tilde{x}_i - \tilde{x}_k}{\|\tilde{x}_i - \tilde{x}_k\|_2}.
\end{align}
% \end{equation}
Here $g_{ik}^r$ is the learnable gate weight controlling the magnitude of coordinate update between residue $r_i$ and $r_k$, computed by a Gated Linear Unit (GLU)~\cite{dauphin2017language}.
% where the message is computed over spatial neighbors within a threshold distance to preserve local structural relevance. \boldsymbol{x}_i^{rr} in Equation~\ref{eq:gate}

% \begin{equation}
% \begin{aligned}
%     % \mathbf{m}_i^{r} &= \sum_{k \in \mathcal{N}(i)} \alpha_{ik}^{r} \mathbf{v}_k^{r}\\
%     g_{ik}^r &= \text{gate} \left( \mathbf{q}_i^r ; \mathbf{k}_k^r ; \phi_{\text{rbf}} \left(\|\tilde{x}_i - \tilde{x}_k\|\right) \right) \\
%     \Delta x_i^{rr} &= \sum_{k \in \mathcal{N}(i)} \alpha_{ik}^{r} \cdot g_{ik}^r \cdot \frac{\tilde{x}_i - \tilde{x}_k}{\|\tilde{x}_i - \tilde{x}_k\|_2}
% \end{aligned}
% \end{equation}
% The message is computed over spatial neighbors within a threshold distance to preserve local structural relevance.

\paragraph{(II) Residue–Atom Attention.}  % \mathbf{Z} =  \sum_{k=1}^{N} molecular
To model the interaction between residues and substrate atoms, we perform cross-modal attention between them. 
% 首先我们通过SNE层获得保持SE(3)等变性的底物原子特征 maintaining the representation and obtain them at the $l$-th layer the residue and substrate atom representations
% 每一行代表一个原子的类型特征（元素类型、芳香性、连接度、氢原子数和杂化类型） \cite{guan2023d}   the atom-level chemical features at \mathbb{R}^{5} $a_j$ is  
%  by the SNE layer
% After $L_e$ layers of SNE, the substrate atom representation $\mathbf{z}$ with SE(3)-equivariance is obtained.   $j$-th
% 
% The attention weight between the residue $r_i$ and the substrate atom $v_j$ has two directions: The residue-substrate attention $\alpha_{ij}^{rs}$ and substrate-residue attention $\alpha_{ij}^{sr}$. They are computed as:
The attention weights $\beta_{ij}^{rs}$ between the residue $r_i$ and the substrate atom $v_j$ are computed as:
% $\{\mathbf{z}_j\}_{j=1}^{M}$ with coordinates $\{y_j\}_{j=1}^{M}$  
% $\left(\|\tilde{x}_i - y_j\|_2\right)$ $\mathbf{q}_i^{s} = \mathbf{W}_Q^{s}\mathbf{h}_i$, 
\begin{equation}
\begin{aligned}
\beta_{ij}^{rs} &= \text{softmax}_i\!\left(\frac{(\mathbf{q}_i^{r})^\top \mathbf{k}_j^{s}}{\sqrt{d}} + \mathbf{W}_{s}^\top \phi_{\text{rbf}}\!\left(\|\tilde{x}_i - y_j\|_2\right)\right),
% \\
% \beta_{ij}^{sr} &= \text{softmax}_j\!\left(\frac{(\mathbf{q}_i^{r})^\top \mathbf{k}_j^{s}}{\sqrt{d}} + \mathbf{W}_{s}^\top \phi_{\text{rbf}}\!\left(\|\tilde{x}_i - y_j\|_2\right)\right)
\label{eq:alpha}
\end{aligned}
\end{equation}
where $\mathbf{k}_j^{s} = \mathbf{W}_K^{s} \cdot \text{LN}(\mathbf{z}_j)$, $\mathbf{v}_j^{s} = \mathbf{W}_V^{s} \cdot \text{LN}(\mathbf{z}_j)$, $\text{softmax}_i$ means normalizing over residues. 

We perform the $\text{softmax}_j$ to the same attention score to obtain the attention weight $\beta_{ij}^{sr}$ over atoms.
% $\alpha_{ij}^{rs}$ is normalized over residues, and $\alpha_{ij}^{sr}$ over substrate atoms. 
% 
% atoms interaction
% Then 
The residue message $\mathbf{m}_i^{rs}$ and substrate atom message $\mathbf{m}_j^{sr}$ is then aggregated as:
% \begin{equation}
\begin{align}
\mathbf{m}_i^{rs} =  \sum_{j \in \mathcal{N}(i)}  \beta_{ij}^{rs} \mathbf{v}_j^{s},\\
% \quad \text{and} \quad
\mathbf{m}_j^{sr} =  \sum_{i \in \mathcal{N}(j)}  \beta_{ij}^{sr} \mathbf{v}_j^{r}.
\end{align}
% \end{equation} 四元组

\paragraph{Feature and Coordinate Updates.}  
% intra-residue and residue-atom After performing the intra-residue and residue-atom attentions,
Next, we update the representations and coordinates of pocket residue and substrate atom based on the obtained messages and attention weights. 
For accurately modeling pocket-substrate interactions, we only update the residue of the pocket sites $\mathcal{\tilde{P}} = \{i \mid \min_{v_j \in \mathcal{V}} \|\tilde{x}_i - y_j\|_2 < d\}$ and all atom of substrate. 
% use the obtained intra-residue and residue–substrate messages to obtain the updated pocket residue representation and coordinate  为了准确模拟口袋  
% 
Specifically, we integrate intra-residue and residue–substrate messages to obtained the updated representations $\mathbf{h}_i'$ and $\mathbf{z}_i'$:
% $\mathbf{h}_i'$, as well as the updated substrate atom coordinate $\mathbf{z}_i'$:
\begin{align}
\mathbf{h}_i' &= \begin{cases}
    \mathbf{h}_i + \text{FFN} \left(m_i^r; m_i^{rs}\right), & i \in \mathcal{\overline{M}} \cap \mathcal{\tilde{P}},
    \\
    \mathbf{h}_i, & \text{otherwise},\end{cases} \\
\mathbf{z}_i' &= \mathbf{z}_i + \text{FFN} \left(m_i^{sr}\right).
\label{eq:feature_update}
\end{align}
Here $\mathcal{\overline{M}}$ is the non-functionally conserved sites.
% $\mathcal{\tilde{P}} = \{i \mid \min_{v_j \in \mathcal{V}} \|\tilde{x}_i - y_j\|_2 < d\}$ is the pocket residue sites. quadruples coordinate update

Then, we use two GLUs to respectively compute the gate weight $g_{ij}^{rs}$ and $g_{ij}^{sr}$ of residues and atoms based on Equation~\ref{eq:gate} and $\mathbf{q}_i^r, \mathbf{k}_j^s,\tilde{x}_i,y_j$.
% and obtain the coordinate update amount $\Delta x_i^{sr}$ for residue $r_i$ and $\Delta y_j^{rs}$ for atom $v_j$:
The coordinate update amount $\Delta x_i^{sr}$ for residue $r_i$ and $\Delta y_j^{rs}$ for atom $v_j$ are computed as:
% to obtain the:
% \begin{equation}
\begin{align}
    % g_{ij}^{rs} &= \text{GLU} \left( \mathbf{q}_i^r ; \mathbf{k}_j^s ; \phi_{\text{rbf}} \left(\|\tilde{x}_i - y_j\|_2\right) \right) \\
    % g_{ij}^{sr} &= \text{GLU} \left( \mathbf{q}_i^r ; \mathbf{k}_j^s ; \phi_{\text{rbf}} \left(\|\tilde{x}_i - y_j\|_2\right) \right) \\ 
    \Delta x_i^{rs} &= \sum_{j \in \mathcal{N}(i)} \beta_{ij}^{rs} \cdot g_{ij}^{rs} \cdot \frac{y_j - \tilde{x}_i}{\|\tilde{x}_i - y_j\|_2}, \\
    \Delta y_j^{sr} &= \sum_{i \in \mathcal{N}(j)} \beta_{ij}^{sr} \cdot g_{ij}^{sr} \cdot \frac{\tilde{x}_i - y_j}{\|\tilde{x}_i - y_j\|_2}.
\end{align}
Finally, coordinates of residues and atoms is updated as $\mathbf{h}_i'$ and $\mathbf{z}_i'$, respectively:
\begin{align}
% \mathbf{h}_i' &= \begin{cases}
%     \mathbf{h}_i + \text{FFN} \left(m_i^r; m_i^{rs}\right), & i \in \mathcal{\overline{M}} \cap \mathcal{\tilde{P}}
%     \\
%     \mathbf{h}_i, & \text{otherwise}\end{cases} \\
% \mathbf{z}_i' &= \mathbf{z}_i + \text{FFN} \left(m_i^{sr}\right)\\
\tilde{x}_i' &= \begin{cases}
    \tilde{x}_i + \Delta x_i^{rr} + \Delta x_i^{rs}, & i \in \mathcal{\overline{M}} \cap \mathcal{\tilde{P}},
    \\
    \tilde{x}_i, & \text{otherwise},\end{cases} \\ 
y_j' &= y_j + \Delta y_j^{sr}. 
\label{eq:feature_update_y}  
\end{align}
We stack $L_{p}$ of RBA modules to obtain the final representation $\hat{\mathbf{h}_i}$ and coordinates $\hat{x_i}$ and $\hat{y_j}$, a prediction head are then employed to predict amino acid type:
% To predict amino acid type, we project the representation into the amino acid vocabulary space: project the representation into the amino acid vocabulary space
\begin{equation}
\hat{s_i} = 
\text{softmax}\!\left(\mathbf{W}_{\text{aa}}\hat{\mathbf{h}_i} + \mathbf{b}_{\text{aa}}\right), 
\label{eq:aa_prediction}
\end{equation}
where $\mathbf{W}_{\text{aa}} \in \mathbb{R}^{|\mathcal{A}| \times h_d}$ and $\mathbf{b}_{\text{aa}} \in \mathbb{R}^{|\mathcal{A}|}$ are trainable parameters, and $|\mathcal{A}| = 20$ denotes the amino acid vocabulary size.   

\subsection{Training Objective}
% pocket coordinate refinement, 
% \begin{equation}
% \mathcal{L}_{\text{seq}} = \frac{1}{|\mathcal{M}_{\text{mask}}|} \sum_{i \in \mathcal{M}_{\text{mask}}} l_{\text{CE}}(s_i^{\text{gt}}, \hat{s}_i)
% \label{eq:seq_loss_pba}
% \end{equation}
% \begin{equation}
% \mathcal{L}_{\text{coord}} = \frac{1}{|\mathcal{M}_{\text{mask}}|} \sum_{i \in \mathcal{M}_{\text{mask}}} l_{\text{Huber}}(x_i^{\text{gt}}, \hat{x}_i)
% \end{equation}
% \begin{equation}
% \mathcal{L}_{\text{pcoord}} = \frac{1}{|\mathcal{M}_{\text{pocket}}|} \sum_{i \in \mathcal{M}_{\text{pocket}}} l_{\text{Huber}}(x_i^{\text{gt}}, \hat{x}_i)
% \label{eq:coord_loss_pba}
% \end{equation}
% \begin{equation}
% \mathcal{L}_{\text{bind}} = \frac{1}{|\mathcal{M}_{\text{pocket}}|} \sum_{i \in \mathcal{M}_{\text{pocket}}} w_p \cdot l_{\text{CE}}(\mathbf{M}_i^{\text{gt}},\hat{\mathbf{M}_i})
% \label{eq:pocket_loss_pba}
% \end{equation}
% \begin{equation}
% \mathcal{L} = \lambda_{\text{seq}} \mathcal{L}_{\text{seq}} + \lambda_{\text{coord}} \mathcal{L}_{\text{coord}} + \lambda_{\text{pcoord}} \mathcal{L}_{\text{pcoord}} + \lambda_{\text{bind}} \mathcal{L}_{\text{bind}},
% \label{eq:total_loss_pba}
% \end{equation}
% \mname\ is trained jointly using tasks of masked enzyme sequence reconstruction, masked coordinate regression, and pocket coordinate refinement. multi-scale structural alignment of masked residues, 
% The training loss $\mathcal{L}$ is a weighted sum of the following losses: masked enzyme 
\mname\ is trained jointly using sequence reconstruction and structural refinement.
% of substrate-binding pockets, and substrate coordinate consistency.  
% 优化目标
The training loss \(\mathcal{L}\) considers the joint optimization objective, including the enzyme sequence loss $\mathcal{L}_{\text{s}}$, pocket loss $\mathcal{L}_{\text{ps}}$, enzyme coordinates loss $\mathcal{L}_{\text{c}}$, pocket coordinates loss $\mathcal{L}_{\text{pc}}$, and substrate coordinates loss $\mathcal{L}_{\text{sub}}$:
% from the sequence and structure of the residue and pocket, and substrate structure, formulated as a weighted sum of the following losses:
% \begin{equation}
\begin{align}
\mathcal{L}_{\text{s}} &= \frac{1}{|\mathcal{\overline{M}}|} \sum_{i \in \mathcal{\overline{M}}} l_{\text{CE}}(s_i^{\text{gt}}, \hat{s}_i), \\
\mathcal{L}_{\text{ps}} &= \frac{1}{|\mathcal{P}|} \sum_{i \in \mathcal{P}} l_{\text{CE}}(s_i^{\text{gt}}, \hat{s}_i), \\
\mathcal{L}_{\text{c}} &= \frac{1}{|\mathcal{\overline{M}}|} \sum_{i \in \mathcal{\overline{M}}} l_{\text{Huber}}(x_i^{\text{gt}}, \hat{x}_i),\\
\mathcal{L}_{\text{pc}} &= \frac{1}{|\mathcal{P}|} \sum_{i \in \mathcal{P}} l_{\text{Huber}}(x_i^{\text{gt}}, \hat{x}_i),\\
\mathcal{L}_{\text{sub}} &= \frac{1}{|M|} \sum_{j} l_{\text{Huber}}(y_j^{\text{gt}}, \hat{y}_j). 
\label{eq:total_loss_pba}
\end{align}
% \end{equation}
Here $\mathcal{P}$ is the ground-truth of pocket residue sites. 
$s_i^{\text{gt}}$ denotes the ground-truth amino acid type, $x_i^{\text{gt}}$ and $y_j^{\text{gt}}$ are correspond to real 3D coordinates of residue $r_i$ and substrate atom $v_j$. 
% the superscript $\text{gt}$ denotes ground-truth values, where $\mathcal{\overline{M}}$ is the set of masked residues during training (i.e., residues at non-functionally conserved sites) and  the index set of 
% (with $M_i^{\text{gt}} \in \{0,1\}$ and 1 indicating a pocket residue);    $\mathbf{M}_i^{\text{gt}}$
% $\hat{s}_i$ and $\hat{x}_i$ are the predicted amino acid distribution and 3D coordinates of residue $i$, $s_{i,g} \in [0,1]$ is the learned pocket importance gate, and $w_i$ is the class weight for addressing the imbalance between pocket and non-pocket residues;   (smooth L1 loss with $\beta=1.0$) $l_{\text{BCE}}$
% and $w_p$ is the classification weight for balancing the amount of pocket and non-pocket residues. , and pocket label   substrate-binding
$l_{\text{CE}}$ and $l_{\text{Huber}}$ denote cross-entropy and Huber loss, respectively. 
Then we obtain the total loss:
\begin{equation}
    \mathcal{L} = \lambda_{\text{s}} \mathcal{L}_{\text{s}} + \lambda_{\text{ps}}\mathcal{L}_{\text{ps}} + \lambda_{\text{c}}\mathcal{L}_{\text{c}} + \lambda_{\text{pc}}\mathcal{L}_{\text{pc}} + \lambda_{\text{sub}}\mathcal{L}_{\text{sub}}, 
\end{equation}
where $\lambda_{[\cdot]}$ are hyperparameters for balancing these losses.
% $\lambda_{\text{s}}$, $\lambda_{\text{ps}}$, $\lambda_{\text{c}}$, $\lambda_{\text{pc}}$ and $\lambda_{\text{sub}}$ 
% 这里应该讲述超参数如何确定。
% By minimizing the total loss in Eq.~\ref{eq:total_loss_pba}, \mname\ learns to generate enzyme sequences that are both functionally coherent and geometrically compatible with their substrate-binding pockets, enabling controllable enzyme generation guided by structural and catalytic priors.

\section{Experiments}

\subsection{EnzyPock: Dataset Curation}

\begin{table}[htbp]
  \caption{Statistics of EnzyPock. Sample size indicates the number of enzyme-substrate pairs. The middle two rows are the number of third and fourth-level EC families. }
  \centering
  \scalebox{.75}{% 0.85倍页面宽度
  \begin{tabular}{lccc} % 第一列左对齐，后三列居中
    \toprule
    Dataset & Training & Validation & Test \\
    \midrule
    Sample Size & 83,062 & 483 & 791 \\
    Third-Level Family & 161 & 18 & 24 \\
    Fourth-Level Family & 973 & 30 & 33 \\
    Average Seq. Length & 324 & 314 & 322 \\
    \bottomrule
  \end{tabular}
    }
  \label{tab:dataset} % 用于交叉引用的标签
\end{table}
% 样本数 for different approaches

\begin{table}[t]
\centering
\caption{
Quantitative results on EnzyPock. The best and second-best results are marked in \textbf{bold} and \underline{underlined}, respectively. 
}
\label{tab:mainresults}
\scalebox{.65}{% 0.85倍页面宽度
\begin{tabular}{lccccc}
\toprule
\textbf{Method} & \textbf{AAR} $\uparrow$  & \textbf{Vina score} $\downarrow$ & \textbf{pLDDT} $\uparrow$ & \textbf{scRMSD} $\downarrow$ & \textbf{scTM} $\uparrow$  \\
\midrule
ESM3 (1.4B) & \underline{0.76} & -3.46 & \textbf{77.87} & \underline{5.26} & \underline{0.89} \\
RFdiffusion2+IF (\textcolor{black}{84M}) & 0.56 & {-6.65} & 68.70 & 6.36 & 0.86 \\
EnzyControl (\textcolor{black}{21M}) & 0.49 & \underline{-6.90} & 58.74 & 8.54 & 0.77 \\
EnzyGen (714M) & 0.64 & {-6.65} & 60.46 & 9.64 & 0.75 \\
\midrule
{\mname\ (798M)} & \textbf{0.77} & \textbf{-7.12} & \underline{74.87} & \textbf{3.10} & \textbf{0.91} \\ 
\midrule
w/o intra-residue attention & 0.73 & -7.08 & 73.63 & 4.54 & 0.88 \\
w/o residue-atom attention & 0.73 & -7.13 & 72.48 & 4.55 & 0.87 \\
w/o pocket \& substrate loss & 0.73 & -7.13 & 74.43 & 3.54 & 0.89 \\
Freezing RFF & 0.70 & -6.82 & 69.75 & 6.16 & 0.84 \\
\bottomrule
\end{tabular}
}
\end{table}

% \begin{table*}[t]
% \centering
% \caption{\textbf{Ablation Study.} Effect of removing key components from \mname.
% Each ablated variant is trained under identical settings. Removing the RBA substantially degrades the performance. \textcolor{red}{can we merge it into table 2? } }
% \label{tab:ablation}
% \begin{tabular}{lccccc}
% \toprule
% \textbf{Method} & \textbf{AAR} $\uparrow$  & \textbf{Vina Score (kcal/mol)} $\downarrow$ & \textbf{pLDDT} $\uparrow$ & \textbf{scRMSD (\AA)} $\downarrow$ & \textbf{scTM} $\uparrow$  \\
% \midrule
% \mname\ & \textbf{--} & \textbf{--} & \textbf{--} & \textbf{--} \\
% w/o intra-residue attention & -- & -- & -- & -- & -- \\
% w/o residue-atom attention & -- & -- & -- & -- & -- \\
% w/o RFF & -- & -- & -- & -- & -- \\
% w/o pocket \& substrate loss & -- & -- & -- & -- & -- \\
% % w/o SNE & -- & -- & -- & -- \\
% % w/o  & -- & -- & -- & -- \\
% % w/o Substrate Condition & -- & -- & -- & -- & -- \\
% \bottomrule
% \end{tabular}
% \end{table*}

\begin{table}[!htbp]
\centering
\caption{
Quantitative results on EnzyBench. 
}
\label{tab:enzybench}
\scalebox{.65}{% 0.85倍页面宽度
\begin{tabular}{lccccc}
\toprule
\textbf{Method} & \textbf{AAR} $\uparrow$  & \textbf{Vina score} $\downarrow$ & \textbf{pLDDT} $\uparrow$ & \textbf{scRMSD} $\downarrow$ & \textbf{scTM} $\uparrow$  \\
\midrule
ESM3 (1.4B) & 0.64 &  -6.58 & \textbf{71.64} & 13.89 & 0.72 \\
RFdiffusion2+IF (\textcolor{black}{84M}) & 0.32 & {-7.15} & 61.91 & 11.81 & 0.72 \\
EnzyControl (\textcolor{black}{21M}) & 0.35 & -7.18 & 60.64 & 10.88 & 0.71 \\
% DiMA (740M) & -- & -- & -- & -- & -- \\
% DPLM (650M) & -- & -- & -- & -- & -- \\
EnzyGen (714M) & \textbf{0.88} & \textbf{-8.27} & \underline{70.10} & \textbf{6.37} & \textbf{0.82} \\
\midrule
% \textbf{\mname-1.5B (ours)} & \textbf{--} & \textbf{--} & \textbf{--} & \textbf{--} & \textbf{--} \\
% \textbf{\mname-798M (ours)} & \textbf{--} & \textbf{--} & \textbf{--} & \textbf{--} & \textbf{--} \\
{\mname\ (798M)} & \underline{0.71} & \underline{-8.17} & {67.77} & \underline{10.48} & \underline{0.72} \\
% w/o intra-residue attention & -- & -- & -- & -- & -- \\
% w/o residue-atom attention & -- & -- & -- & -- & -- \\
% w/o RFF & -- & -- & -- & -- & -- \\
% w/o pocket \& substrate loss & -- & -- & -- & -- & -- \\
\bottomrule
\end{tabular}
}
\end{table}

\begin{table*}[t]
\centering
\caption{Binding affinity (Vina score) and Foldability (pLDDT) comparison across 9 main EC-2 families on EnzyPock.}
\label{tab:ec_combined}
\vspace{-2mm}
\scalebox{.75}{
\begin{tabular}{l *{9}{c} | c}
\toprule
\multicolumn{11}{c}{\textbf{Binding affinity (Vina score $\downarrow$)}} \\
\midrule
EC Family 
& 1.1 & 2.1 & 2.3 & 2.7 & 3.2 & 3.5 & 4.1 & 5.2 & 5.3 & Avg. \\
\midrule
ESM3 (1.4B) 
& -4.55 & -0.60 & \textbf{-0.00} & -2.04 & -1.53 & -5.67 & -0.91 & -3.71 & -0.14 & -3.52 \\
RFdiffusion2+IF (84M)
& \textbf{-5.62} & \textbf{-6.38} & -9.14 & \textbf{-8.57} & \textbf{-7.27} & -6.47 & -6.88 & -6.79 & -5.30 & -7.01 \\
EnzyControl (21M)
& -4.74 & -6.28 & -8.57 & -8.48 & -6.59 & -6.45 & -6.96 & \textbf{-7.10} & -5.88 & -6.93 \\
EnzyGen (714M)
& -4.75 & -6.12 & -9.03 & -8.44 & -6.36 & -5.96 & -6.97 & -6.80 & \textbf{-6.06} & -6.66 \\
\midrule
\textbf{\mname\ (798M)}
& -4.89 & -6.21 & \textbf{-9.37} & -8.39 & -6.97 & \textbf{-6.85} & \textbf{-7.15} & -6.98 & -5.84 & \textbf{-7.14} \\
\midrule
\multicolumn{11}{c}{\textbf{Foldability (pLDDT $\uparrow$)}} \\
\midrule
EC Family 
& 1.1 & 2.1 & 2.3 & 2.7 & 3.2 & 3.5 & 4.1 & 5.2 & 5.3 & Avg. \\
\midrule
ESM3 (1.4B) 
& \textbf{83.41} & \textbf{80.82} & \textbf{82.21} & \textbf{75.80} & 74.84 & \textbf{80.37} & \textbf{67.18} & 78.89 & 84.58 & \textbf{77.91} \\
RFdiffusion2+IF (84M)
& 77.15 & 77.90 & 70.07 & 73.15 & 43.66 & 71.84 & 56.38 & 73.07 & 81.63 & 68.56 \\
EnzyControl (21M)
& 68.59 & 76.06 & 71.05 & 70.58 & 37.74 & 55.16 & 50.05 & 62.83 & 82.42 & 58.62 \\
EnzyGen (714M)
& 63.31 & 75.78 & 53.34 & 70.47 & 45.00 & 58.73 & 50.65 & 63.08 & 76.60 & 60.33 \\
\midrule
\textbf{\mname\ (798M)}
& 79.96 & 76.27 & 75.68 & 71.34 & \textbf{75.59} & 79.25 & 49.68 & \textbf{79.38} & \textbf{85.34} & 74.89 \\
\bottomrule
\end{tabular}
}
\end{table*}

\begin{figure*}[t]
\centering
\includegraphics[width=0.85\linewidth]{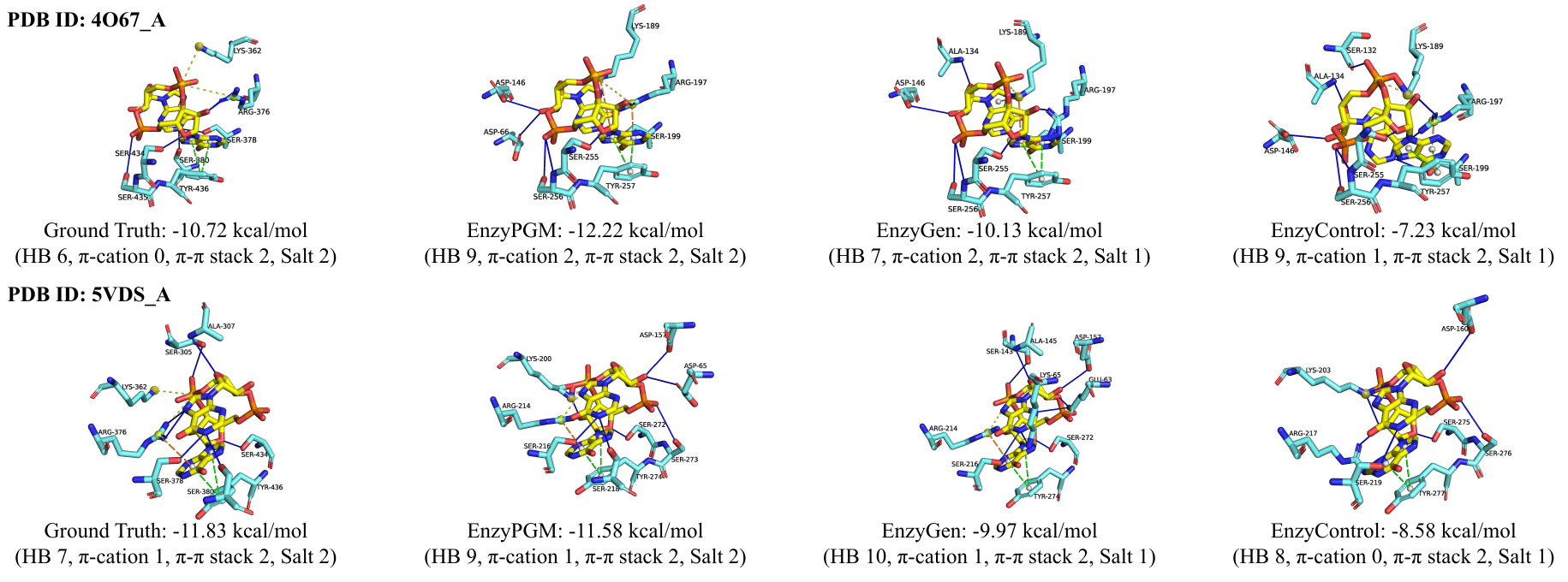}
\vspace{-2mm}
\caption{\textbf{Case Study.}
Visualization of ground-truth, \mname-generated, and 2 baseline-generated enzyme-substrate complexes in order. Each panel displays the enzyme pocket (blue) and substrate (yellow), alongside key metrics: the Vina score (binding affinity) and the number of enzyme-substrate interactions: hydrogen bond (HB), $\pi$-cation, $\pi$-$\pi$ stack, and salt bridge (Salt).
}
\label{fig:pocketviz}
\end{figure*}
% 依次

\begin{figure*}[htp!]
\centering
\begin{tabular}{lccc}
\includegraphics[width=0.18\linewidth]{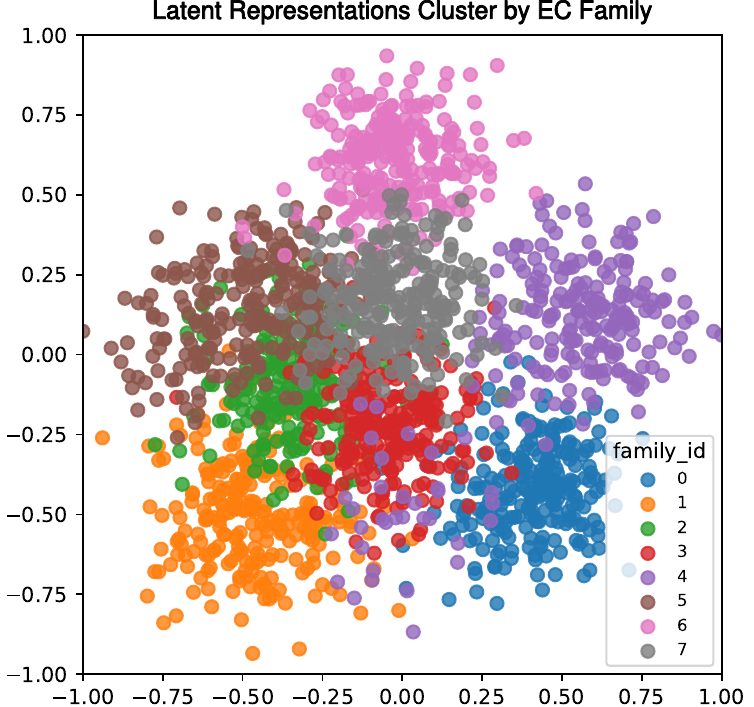}&\includegraphics[width=0.18\linewidth]{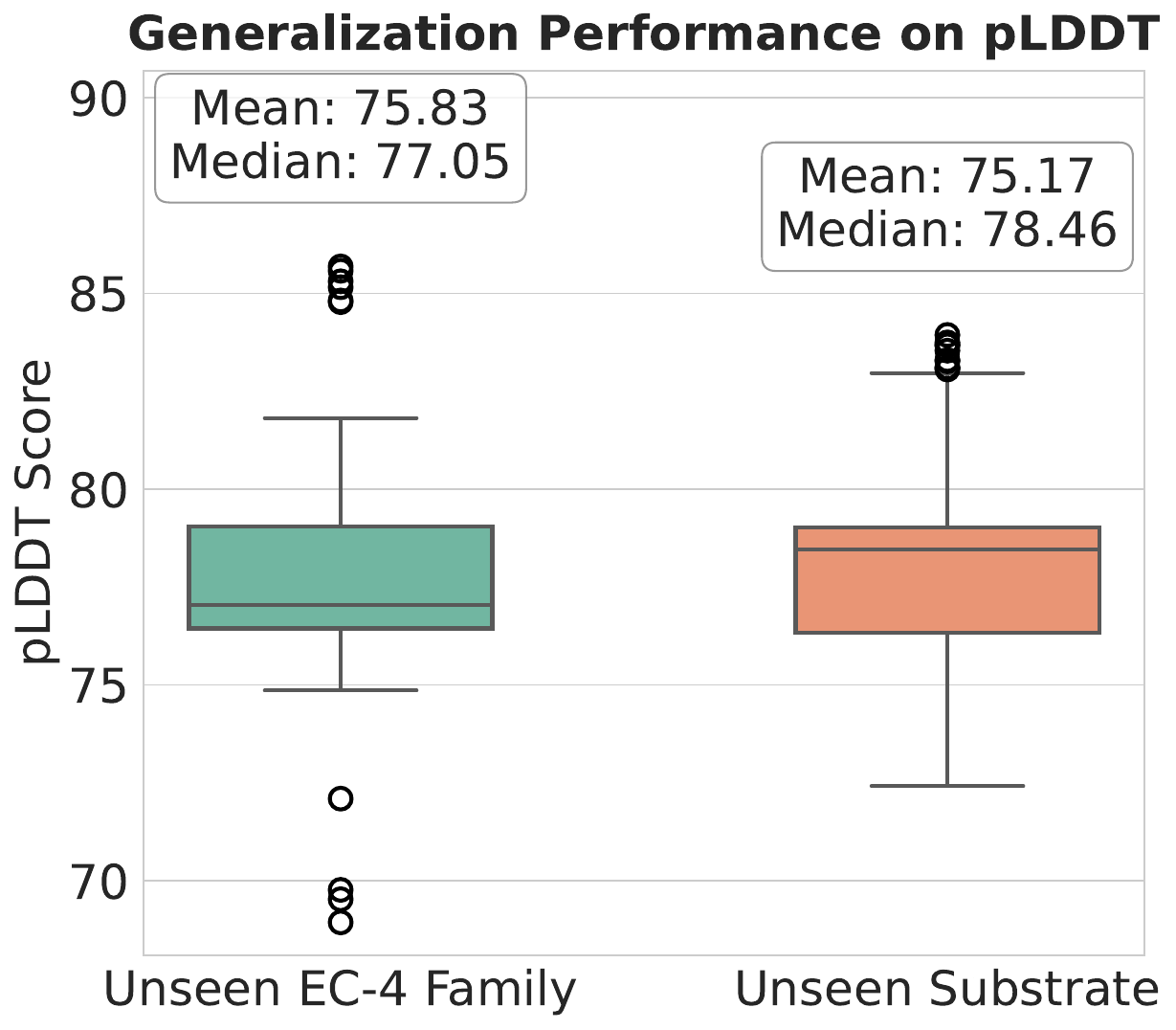}&\includegraphics[width=0.20\linewidth]{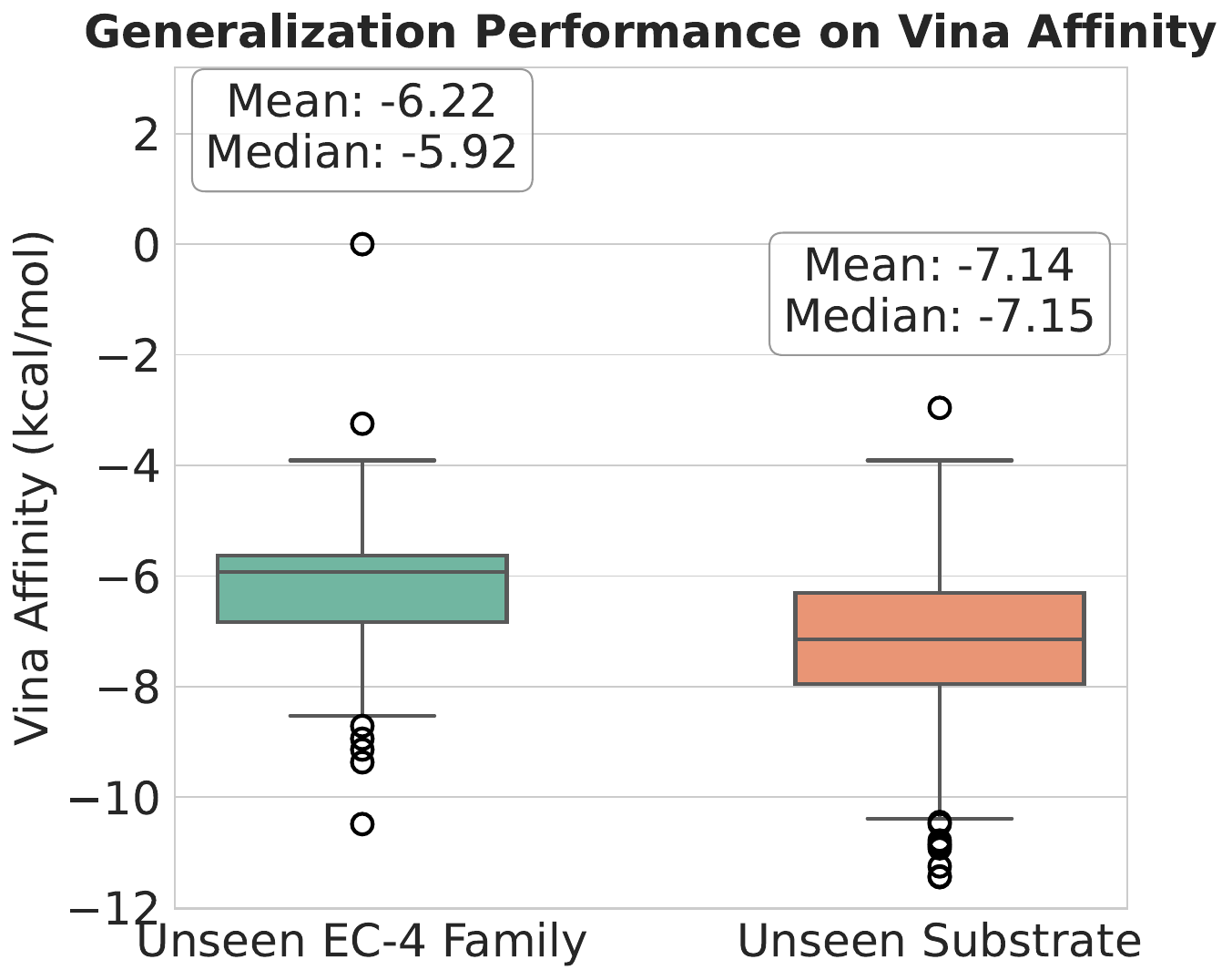}&\includegraphics[width=0.24\linewidth]{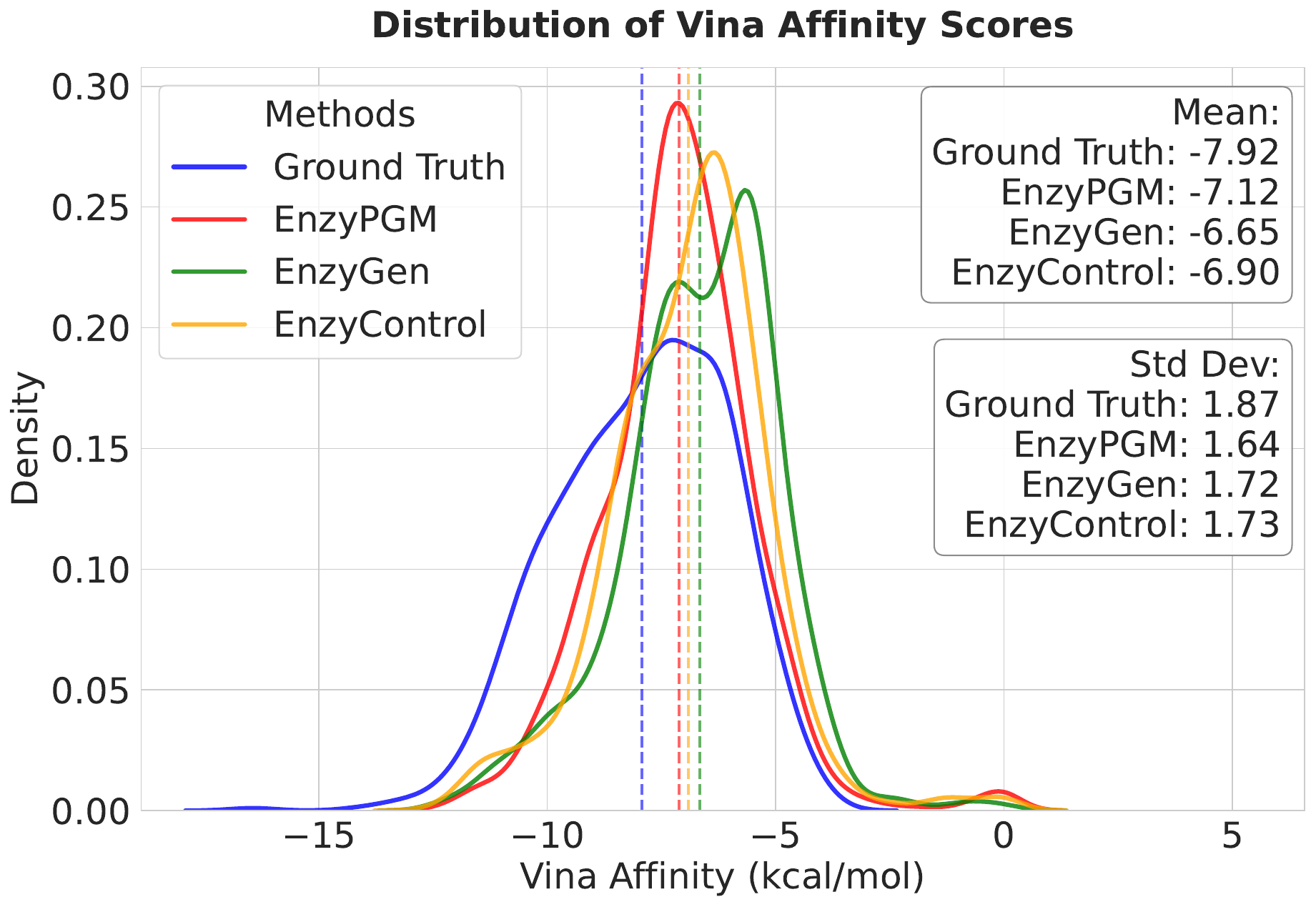}\\
\tiny(a) Visualization on representation learning &\tiny(b) Generalization capability on sequence validity&\tiny(c) Generalization capability on binding affinity&\tiny(d) Comparison on affinity distribution\\ 
\end{tabular}
\caption{The different analysis on \mname. \label{fig:analysis}}
\end{figure*}
% \label{fig:analysis-a} 
% \begin{figure}[htp!]
%     \centering
%     \includegraphics[width=1.0\linewidth]{figures/case_study.pdf}
%     \caption{
%  The case study on the different approaches.
%     } Analysis of 
% \end{figure}

To construct a unified enzyme-pocket dataset that includes enzymes, substrates, EC numbers, and substrate-binding pockets, we employ a data collection and processing pipeline.
% ~\cite{francoeurThreeDimensionalConvolutionalNeural2020} 数据收集和处理流程 Data collection and processing procedures (Figure \ref{append:data_collect})
% 首先，我们先收集PDB和
% 接下来，其次  
First, we collect the protein-ligand pairs on PDBbind and CrossDocked.
Next, enzymes in these samples are identified via PDB\footnote{\url{https://www.rcsb.org/}}-to-UniProt\footnote{\url{https://www.uniprot.org/}} ID mapping and EC number annotation from UniProt.
% Specifically, we first retrieve the EC number for each enzyme in our collected enzyme-pocket dataset from the Uniprot\footnote{\url{https://www.uniprot.org/}} database.
% We then classify all enzymes corresponding to their EC number and perform MSA on these enzymes of fourth-level EC families using the Muscle5 tool~\cite{edgar2022muscle5} to identify conserved sites for each family. classify all enzymes corresponding to their EC number and
We then perform MSA on these enzymes according to their fourth-level EC families using the Muscle5 tool~\cite{edgar2022muscle5} to identify conserved sites for each family.
In practice, we select residues present at the same position in more than 30\% sequences as the conserved sites.
% Pockets are defined as residues within 10 \,\AA of the substrate, with consistent distance alignment across datasets.   
Finally, we extract the substrate-binding pockets based on Equation \ref{eq:pocketdef}. 

% 我们在处理这些酶的时候发现许多酶是由多条链构成的，如果将这些链全部合并在一起处理会降低同源MSA分析的效果，同时序列长度过长还会提高训练开销。为此，我们对多链蛋白仅保留第一条EC注释链，以避免同源MSA分析性能损失并减少不必要的训练成本，并且通过CHEBI和KEGG补充底物信息以确保完整性。
% Multi-chain proteins retain only the first EC-annotated chain to avoid loss of homologous MSA analysis performance and reduce unnecessary training costs, and substrate information is complemented via CHEBI and KEGG to ensure completeness.   
In this process, we observe that many enzymes consist of multiple chains, and merging all chains for processing degrades the quality of homologous MSA analysis. Concatenating these chains leads to excessively long sequences and increases training cost. 
Therefore, we retain only the first EC-annotated chain for multi-chain proteins.
% to avoid MSA performance loss and reduce unnecessary computation. 
% Substrate information is further supplemented using CHEBI and KEGG to ensure completeness.

% since PDBBind is a dataset verified by biological wetness experiments, which is more realistic and convincing. 
On the other hand, we adopt them according to the priority order of PDBBind and CrossDocked when records from different datasets conflict because PDBBind is validated by biological wet-lab experiments and CrossDocked is mainly composed of model predictions, which makes the former more reliable.
% PDBBind是经过生物湿实验验证的数据集，更真实且更有说服力,而EnzyBench相较于CrossDocked的优势在于，CrossDocked是模型预测的结合结果，而EnzyBench是直接从Brenda中筛选的数据。更可靠的 (>70\% identity)  the dataset
To prevent data leakage, all {17,404} PDB entries in EnzyPock are clustered with a sequence identity threshold of 70\%. We split these samples of different clusters into training, validation, and test sets with non-overlapping clusters. The details of the dataset are shown in Table~\ref{tab:dataset}.  
% }\textbf{17,404} PDB entries 
% Each sample includes enzyme sequence/structure, substrate molecular graph, and pocket geometry ().
% To prevent data leakage, we cluster the PDB entries with a sequence identity threshold of 50%. We ensure that no PDB entries in the validation or test sets belong to the same cluster as those in the training set. 

\subsection{Experimental Settings}

\noindent\textbf{Baselines}: 
(1) ESM3~\cite{Lin2024} is an advanced multimodal protein language model generating the sequence, structure, and function of proteins;
(2) RFDiffusion2+IF first uses RFDiffusion2~\cite{ahern2025atom} to design enzyme structures based on given functionally important sites, then applies ProteinMPNN to generate sequences from the predicted structures; % ~\cite{Dauparas2022ProteinMPNN}
(3) EnzyControl~\cite{songenzycontrol} is a diffusion model for enzyme backbone generation based on functional and substrate-specific control.
(4) EnzyGen~\cite{EnzyGen2024} is a substrate-specified enzyme generative model based on the given functionally conserved sites and EC label.
Since ESM3 and RFDiffusion2 do not release their training script, we directly use their released model weight to generate the sequences.
For EnzyControl, we use the public weight and follow their evaluation pipeline to test performance because we encounter challenges in using their code to train.
% Furthermore, we use the public weight of EnzyControl to test its performance since we enzyme  on our dataset and EnzyBench
% EnzyControl的数据转换脚本有问题，我们使用它公布的权重在Enzypock和EnzyBench上推理。 enzyme to infer on the benchmark without training,  ~\cite{Li2024PocketGen_Generalized} without training and inference code and predict their all-atom structures by ESMFold
% it feel good.

\noindent\textbf{Evaluation Metrics}: (1) Amino Acid Recovery (AAR) refers to the percentage of correctly recovered residue types in generated enzymes;
(2) Vina score~\cite{Trott2010Vina} to evaluate the binding affinity between enzymes and substrates using Gnina~\cite{mcnuttGNINA10Molecular2021};
(3) pLDDT~\cite{Jumper2021AlphaFold2} to measure the foldability of generated sequences using ESMFold~\cite{doi:10.1126/science.ade2574}; 
(4) scRMSD~\cite{Trippe2023ICLR} to assess the structural validity of generated backbones;
(5) scTM to evaluate the structural design quality of generated enzymes.
We adopt these metrics on EnzyPock and EnzyBench to test the performance of different approaches. 
% All reported metrics are computed on these ESMFold-predicted structures, ensuring that differences in evaluation outcomes reflect genuine differences in backbone design rather than variations in sequence modeling or structure prediction protocols. and EnzyBench to test the performance of different approaches
To ensure a fair comparison between sequence- and structure-only modeling approaches, we uniformly use ESMFold-predicted structures to evaluate structure-related metrics, thereby eliminating evaluation variations from different modeling ways. 
\noindent\textbf{Implementation Details}: \mname\ is composed of three stacked RFF and RBA modules, where each RFF contains 11 Transformer blocks initialized with ESM2 weights~\cite{Lin2022ESM2}. A total of 3 individual SNE layers are used to process the substrate, and RBA only updates the pocket area within a 10\AA\ radius of the substrate.
The total parameter of \mname\ is 798M. The spatial $K$-nearest neighbor number in RFF, substrate SNE, and RBA are set to 30, 16, and 10, respectively.
% 在实验中我们发现序列重建任务对序列整体生成质量影响较大，而结构优化对于口袋的生成质量影响较大，因此我们在训练开始时设置超参数为xxx,训练后期设置为xxx
The hyperparameters $\lambda_{\text{s}}$, $\lambda_{\text{ps}}$, $\lambda_{\text{c}}$, $\lambda_{\text{pc}}$, and $\lambda_{\text{sub}}$ are set to 1, 0.5, 1, 0.5, and 1, respectively.
%     \mathcal{L} = \lambda_{\text{s}} \mathcal{L}_{\text{s}} + \lambda_{\text{ps}}\mathcal{L}_{\text{ps}} + \lambda_{\text{c}}\mathcal{L}_{\text{c}} + \lambda_{\text{pc}}\mathcal{L}_{\text{pc}} + \lambda_{\text{sub}}\mathcal{L}_{\text{sub}},  8\times10^{-6}
% RBA module updates only the pocket area with a 10\,\AA\ cutoff radius of substrates.
% 批次大小随GPU显存容量动态调整。 
For training \mname\ on EnzyPock, we adopt AdamW with a learning rate of 1e-5 and a cosine learning schedule, dynamic batch size on GPU memory capacity, and use four NVIDIA A100 GPUs to train 30 epochs with 8000 warmup steps.
% Pocket refinement is performed for 3 iterative rounds per forward pass. 
% All experiments are conducted on four NVIDIA A100 GPUs and require approximately 72 hours per full-scale run.

\subsection{Quantitative Results}

\noindent\textbf{\mname\ realizes more effective enzyme design.}
% We compare \mname\ with the enzyme design baselines. As shown in Table~\ref{tab:mainresults}, \mname\ achieves the highest pLDDT, and the lowest Vina score, surpassing EnzyGen by xxx\% and xxx\%, reducing the RMSD and average binding energy by xxx\,kcal/mol and xxx \AA. enzyme design
% We compare \mname\ with the baselines. 
As shown in Table~\ref{tab:mainresults}, \mname\ achieves the highest AAR of 0.77 and the lowest Vina score of -7.12 on EnzyPock, surpassing EnzyGen by 0.13, reducing the average binding energy by 0.47\,kcal/mol.
Meanwhile, \mname\ exhibits state-of-the-art performance in scRMSD and scTM on EnzyPock, suggesting the generated enzymes are well-folded and of high substrate affinity.
Notably, ESM3 obtains the highest pLDDT on EnzyPock, showing that large-scale pre-training promotes the enzyme design learning, while \mname\ achieves a close level through fine-tuning.
Also, we test all approaches on EnzyBench (Table~\ref{tab:enzybench}), confirming that \mname\ has desirable performance and generalization, particularly in binding affinity.
Specifically, we demonstrate the performance of all approaches on 9 main EC-2 families of EnzyPock in Table~\ref{tab:ec_combined}.
These results demonstrate that pocket modeling benefits to generate enzymes with higher substrate-binding capability and structural validity, leading to more effective enzyme design.

\subsection{Ablation Studies}
\paragraph{All designs in \mname\ contribute to enzyme generation.}
We conduct ablation experiments to assess the contribution of core designs in \mname, which are the intra-residue and residue-atom attention, pocket and substrate loss, and RFF module.
As shown in Table~\ref{tab:mainresults}, removing two attentions results in a 1.24 and 2.39 drop in pLDDT, respectively.
% and a 0.8\,\AA\ rise in scRMSD, respectively.  generation
% indicating that explicit modeling of substrate-pocket interactions is  foldability essential for accurate enzyme design. and a 1.24 decrease in pLDDT, modeling of residue–residue and residue–substrate interactions terms of   and foldability
This confirms that the bi-scale attention improves performance in enzyme sequence design.
Removing the pocket and substrate loss leads to a modest drop in overall performance, confirming that they promote model training.
% critical for preserving structural foldability and validity. 
% Removing pocket refinement raises the average binding energy by 0.5\,kcal/mol, revealing that iterative geometric correction enhances spatial alignment and interaction stability. 赋予  promote    
% These results confirm that \mname’s bilevel attention and equivariant architecture jointly incentivize its enzyme design performance. lowers scRMSD by 3.2\AA and scTM by 2 RBA and losses
Freezing the RFF during training results in a performance drop across all metrics, suggesting that the backbone model is essential for \mname.
These results confirm that \mname’s core design jointly promote its enzyme design performance.

\subsection{Case Studies} 
% It designs this.
\paragraph{\mname\ designs high-affinity binding pockets.}
To demonstrate \mname's performance in generating pockets, we perform a case study to compare it with 2 baselines and the ground truth.
Visualization in Figure~\ref{fig:pocketviz} reveals that \mname\ generates valid substrate-binding pockets with a lower Vina score than baselines and more abundant hydrogen bonds, which means the residues in the pocket form strong chemical bonds with the substrate.
% \mname\ yields a Vina score (-12.11 kcal/mol) closer to the ground truth (-8.21 kcal/mol) than EnzyGen (-6.961 kcal/mol), paired with more abundant hydrophobic interactions (14 vs. 4 in ground truth, 6 in EnzyGen)—reflecting its superior ability to recapitulate native-like substrate binding affinity and interaction patterns.  
% catalytic residues forming coherent active-site cavities that accurately follow substrate structure.   
% 其中催化位点的残基与底物之间形成了坚固的结合化学键 不规则的
% 相反，EnzyGen 经常生成不规则的结合区域,这些区域中的残基难以与底物形成稳定化学键。pocket-generation
% 这些结果强调，口袋增强训练不仅提高了全局酶质量，而且增强了催化中心的原子级空间一致性。
In contrast, EnzyGen and EnzyControl generate irregular binding regions where residues struggle to form stable chemical bonds with the substrate. 
These results highlight that pocket-conditioned training not only improves enzyme design quality but also generates valid and stable substrate-binding pockets.

% 深入理解\mname structurally valid 稳定的 robust
\section{Analysis}

To gain deeper insights into \mname, we conduct comprehensive empirical analysis, including representation analysis, generalization evaluations, and binding affinity studies.
\subsection{\mname\ learns EC family-aware representations}
We project 2,000 enzymes into representation space from 8 EC-4 families by \mname\ and use t-SNE~\cite{Maaten2008TSNE} to visualize their latent distribution.
% The resulting clusters align closely with EC classes, while EnzyGen’s representations show less relevance to EC families.  and UMAP McInnes2018UMAP
As shown in Figure~\ref{fig:analysis}(a), enzymes form distinct, well-separated clusters according to different families, which means that enzymes catalyzing similar reactions cluster naturally.
% In contrast, EnzyGen’s latent representations exhibit lower cluster purity (68\% vs. 89\% for \mname), with significant overlap between functionally distinct families.  
% and substrate categories, xiang 相关性
% while EnzyGen’s representations show less functional separation.  biological interpretable
This demonstrates that \mname\ learns function-aware latent representations, which benefits the design of substrate-specific enzymes.
% where enzymes catalyzing similar reactions cluster naturally. 
% Further inspection of RBA attention heads reveals emergent specialization for recognizing hydrogen bonding and $\pi$–$\pi$ stacking motifs, implying that \mname\ automatically internalizes molecular interaction priors relevant to catalysis.  (Tanimoto coefficient)  (20 novel small molecules)

\subsection{\mname\ generalizes on unseen families and novel substrates}
To evaluate generalization, we construct a test set that excludes all EC-4 families and substrates that appear in the training set.
% two test sets: (1) unseen families: enzyme families from EC fourth-level classes entirely excluded during training; (2) novel substrates: substrates with <30\% structural similarity to training substrates. 出现在训练集中的所有家族和底物   only a modest drop compared to the total test set.
As shown in Figure~\ref{fig:analysis}(b) and \ref{fig:analysis}(c), \mname\ achieves an average pLDDT of 75.83 and Vina score of -6.22 on unseen families, 75.17 and -7.14 on unseen substrates.  
% For unseen substrates, it maintains an average pLDDT of 75.17 and a Vina score of -6.94.
The performance is almost consistent with the full test set, which indicates \mname\ can effectively design enzymes with high substrate binding for unseen reactions without retraining.
\subsection{\mname\ designs enzymes with normal affinity distributions}
We demonstrate the distribution of Vina scores obtained by different approaches in the EnzyPock test set.
As shown in Figure~\ref{fig:analysis} (d), the distribution obtained by \mname\ is closer to the ground truth than other approaches, with a closer mean and lowest variance. 
This demonstrates that \mname\ can design normal binding pockets for specific enzymatic reactions.

\section{Conclusion}

In this work, we have proposed \mname, a unified pocket-conditioned generative model for substrate-specific enzyme design. 
Unlike prior approaches, it integrates residue-atom bi-scale attention to model pocket-substrate interactions and fuse enzyme function priors to design enzymes with the binding pocket.
% , jointly designing enzymes and their binding pockets.
To jointly model enzymes and pockets, we also curate EnzyPock, a comprehensive dataset with {84,336} enzyme-substrate pairs across {1,036} fourth-level EC families.
Experiments on EnzyPock show \mname\ outperforms state-of-the-art methods in both binding affinity and structure validity metrics. Further analyses confirm its strong generalization to unseen families and novel substrates, highlighting its effectiveness for substrate-specific enzyme design.
\bibliographystyle{named}
\bibliography{ijcai26}

@article{Bar-Even2011,
  title     = {The moderately efficient enzyme: Evolutionary and physicochemical trends shaping enzyme parameters},
    author = "Arren Bar-Even and Elad Noor and Yonatan Savir and Wolfram Liebermeister and Dan Davidi and Tawfik, \{Dan S.\} and Ron Milo",
    year = "2011",
    month = may,
    day = "31",
    language = "English",
    volume = "50",
    pages = "4402--4410",
    journal = "Biochemistry",
    issn = "0006-2960",
    publisher = "American Chemical Society",
    number = "21",
}

@article{Huang2021,
  title={Controlling the substrate specificity of an enzyme through structural flexibility by varying the salt-bridge density},
  author={Huang, Juan and Xu, Qin and Liu, Zhuo and Jain, Nitin and Tyagi, Madhusudan and Wei, Dong-Qing and Hong, Liang},
  journal={Molecules},
  volume={26},
  number={18},
  pages={5693},
  year={2021},
  publisher={MDPI}
}

@article{Anishchenko2021,
  title={Accurate prediction of protein structures and interactions using a three-track neural network},
  author={Baek, Minkyung and DiMaio, Frank and Anishchenko, Ivan and Dauparas, Justas and Ovchinnikov, Sergey and Lee, Gyu Rie and Wang, Jue and Cong, Qian and Kinch, Lisa N and Schaeffer, R Dustin and others},
  journal={Science},
  volume={373},
  number={6557},
  pages={871--876},
  year={2021},
  publisher={American Association for the Advancement of Science}
}

@article{Madani2022,
  title={Progen2: exploring the boundaries of protein language models},
  author={Nijkamp, Erik and Ruffolo, Jeffrey A and Weinstein, Eli N and Naik, Nikhil and Madani, Ali},
  journal={Cell Systems},
  volume={14},
  number={11},
  pages={968--978},
  year={2023},
  publisher={Elsevier}
}

@article{RFdiffusion2023,
  title={De novo design of protein structure and function with RFdiffusion},
  author={Watson, Joseph L and Juergens, David and Bennett, Nathaniel R and Trippe, Brian L and Yim, Jason and Eisenach, Helen E and Ahern, Woody and Borst, Andrew J and Ragotte, Robert J and Milles, Lukas F and others},
  journal={Nature},
  volume={620},
  number={7976},
  pages={1089--1100},
  year={2023},
  publisher={Nature Publishing Group UK London}
}

@inproceedings{EnzyGen2024,
  title={Generative Enzyme Design Guided by Functionally Important Sites and Small-Molecule Substrates},
  author={Song, Zhenqiao and Zhao, Yunlong and Shi, Wenxian and Jin, Wengong and Yang, Yang and Li, Lei},
  booktitle={ICML},
  pages={46259--46279},
  year={2024},
  organization={PMLR}
}

@article{wang2025artificial,
  title={Artificial intelligence technology assists enzyme prediction and rational design},
  author={Wang, Yuhang and Han, Shuangxin and Wang, Yi and Liang, Quanfeng and Luo, Wei},
  journal={Journal of Agricultural and Food Chemistry},
  volume={73},
  number={12},
  pages={7065--7073},
  year={2025},
  publisher={ACS Publications}
}

@inproceedings{songenzycontrol,
  title={EnzyControl: Adding Functional and Substrate-Specific Control for Enzyme Backbone Generation},
  author={Song, Chao and Liu, Zhiyuan and Huang, Han and Wang, Liang and Wang, Qiong and Shi, Jian-Yu and Yu, Hui and Zhou, Yihang and Zhang, Yang},
  year={2025},
  booktitle={NeurIPS}
}

@article{hua2024reaction,
  title={Reaction-conditioned de novo enzyme design with genzyme},
  author={Hua, Chenqing and Lu, Jiarui and Liu, Yong and Zhang, Odin and Tang, Jian and Ying, Rex and Jin, Wengong and Wolf, Guy and Precup, Doina and Zheng, Shuangjia},
  journal={arXiv preprint arXiv:2411.16694},
  year={2024}
}

@inproceedings{SENZ2025,
  title={Retrieval Augmented Zero-Shot Enzyme Generation for Specified Substrate},
  author={Du, Jiahe and Zhou, Kaixiong and Hong, Xinyu and Xu, Zhaozhuo and Xu, Jinbo and Huang, Xiao},
  year={2025},
  booktitle={ICML}
}

@inproceedings{Stark2022EquiBind,
  title={Equibind: Geometric deep learning for drug binding structure prediction},
  author={St{\"a}rk, Hannes and Ganea, Octavian and Pattanaik, Lagnajit and Barzilay, Regina and Jaakkola, Tommi},
  booktitle={ICML},
  pages={20503--20521},
  year={2022},
  organization={PMLR}
}

@article{Jumper2021AlphaFold2,
  title={Highly accurate protein structure prediction with AlphaFold},
  author={Jumper, John and Evans, Richard and Pritzel, Alexander and Green, Tim and Figurnov, Michael and Ronneberger, Olaf and Tunyasuvunakool, Kathryn and Bates, Russ and {\v{Z}}{\'\i}dek, Augustin and Potapenko, Anna and others},
  journal={Nature},
  volume={596},
  number={7873},
  pages={583--589},
  year={2021},
  publisher={Nature Publishing Group UK London}
}

@book{Lesk2010,
  title={Introduction to protein architecture: the structural biology of proteins},
  author={Lesk, Arthur M},
  volume={741},
  year={2001},
  publisher={Oxford university press Oxford}
}

@article{stank2016protein,
  title={Protein binding pocket dynamics},
  author={Stank, Antonia and Kokh, Daria B and Fuller, Jonathan C and Wade, Rebecca C},
  journal={Accounts of Chemical Research},
  volume={49},
  number={5},
  pages={809--815},
  year={2016},
  publisher={ACS Publications}
}

@article{edgar2022muscle5,
  title={Muscle5: High-accuracy alignment ensembles enable unbiased assessments of sequence homology and phylogeny},
  author={Edgar, Robert C},
  journal={Nature communications},
  volume={13},
  number={1},
  pages={6968},
  year={2022},
  publisher={Nature Publishing Group UK London}
}

@article{Tristem2000,
  title={Predicting functionally important residues from sequence conservation},
  author={Capra, John A and Singh, Mona},
  journal={Bioinformatics},
  volume={23},
  number={15},
  pages={1875--1882},
  year={2007},
  publisher={Oxford University Press}
}

@article{tipton2000history,
  title={History of the enzyme nomenclature system},
  author={Tipton, Keith and Boyce, Sin{\'e}ad},
  journal={Bioinformatics},
  volume={16},
  number={1},
  pages={34--40},
  year={2000},
  publisher={Oxford University Press}
}

@article{Vaswani2017,
  title={Attention is all you need},
  author={Vaswani, Ashish and Shazeer, Noam and Parmar, Niki and Uszkoreit, Jakob and Jones, Llion and Gomez, Aidan N and Kaiser, {\L}ukasz and Polosukhin, Illia},
  journal={NeurIPS},
  volume={30},
  year={2017}
}

@article{Xiong2020,
  title={On layer normalization in the transformer architecture},
  author={Xiong, Ruibin and Yang, Yiming and He, Di and Zheng, Kai and Zheng, Shuxin and Xing, Chenxiang and Zhang, Huishuai and Lan, Yanyan and Wang, Liqiang and Liu, Tie-Yan},
  journal={arXiv preprint arXiv:2002.04745},
  year={2020}
}

@article{Lin2024,
  title={Simulating 500 million years of evolution with a language model},
  author={Hayes, Thomas and Rao, Roshan and Akin, Halil and Sofroniew, Nicholas J and Oktay, Deniz and Lin, Zeming and Verkuil, Robert and Tran, Vincent Q and Deaton, Jonathan and Wiggert, Marius and others},
  journal={Science},
  volume={387},
  number={6736},
  pages={850--858},
  year={2025},
  publisher={American Association for the Advancement of Science}
}

@article{Trott2010Vina,
  title={AutoDock Vina: improving the speed and accuracy of docking with a new scoring function, efficient optimization, and multithreading},
  author={Trott, Oleg and Olson, Arthur J},
  journal={Journal of Computational Chemistry},
  volume={31},
  number={2},
  pages={455--461},
  year={2010},
  publisher={Wiley Online Library}
}

@inproceedings{Trippe2023ICLR,
  title={Diffusion Probabilistic Modeling of Protein Backbones in 3D for the motif-scaffolding problem},
  author={Trippe, Brian L and Yim, Jason and Tischer, Doug and Baker, David and Broderick, Tamara and Barzilay, Regina and Jaakkola, Tommi S},
  year={2023},
  booktitle={ICLR}
}

@article{mcnuttGNINA10Molecular2021,
  title = {{{GNINA}} 1.0: Molecular Docking with Deep Learning},
  shorttitle = {{{GNINA}} 1.0},
  author = {McNutt, Andrew T. and Francoeur, Paul and Aggarwal, Rishal and Masuda, Tomohide and Meli, Rocco and Ragoza, Matthew and Sunseri, Jocelyn and Koes, David Ryan},
  year = 2021,
  month = jun,
  journal = {Journal of Cheminformatics},
  volume = {13},
  number = {1},
  pages = {43},
}

@article{zhangEfficientGenerationProtein2024a,
  title={Efficient generation of protein pockets with PocketGen},
  author={Zhang, Zaixi and Shen, Wan Xiang and Liu, Qi and Zitnik, Marinka},
  journal={Nature Machine Intelligence},
  volume={6},
  number={11},
  pages={1382--1395},
  year={2024},
  publisher={Nature Publishing Group UK London}
}

@article{Li2024PocketGen_Generalized,
  title={Generalized protein pocket generation with prior-informed flow matching},
  author={Zhang, Zaixi and Zitnik, Marinka and Liu, Qi},
  journal={NeurIPS},
  volume={37},
  pages={38559--38589},
  year={2024}
}

@article{Lin2022ESM2,
  title={Language models of protein sequences at the scale of evolution enable accurate structure prediction},
  author={Lin, Zeming and Akin, Halil and Rao, Roshan and Hie, Brian and Zhu, Zhongkai and Lu, Wenting and dos Santos Costa, Allan and Fazel-Zarandi, Maryam and Sercu, Tom and Candido, Sal and others},
  journal={bioRxiv},
  year={2022},
  publisher={Cold Spring Harbor Laboratory}
}

@article{ahern2025atom,
  title={Atom level enzyme active site scaffolding using RFdiffusion2},
  author={Ahern, Woody and Yim, Jason and Tischer, Doug and Salike, Saman and Woodbury, Seth M and Kim, Donghyo and Kalvet, Indrek and Kipnis, Yakov and Coventry, Brian and Altae-Tran, Han Raut and others},
  journal={bioRxiv},
  pages={2025.04.09.648075},
  year={2025},
  publisher={Cold Spring Harbor Laboratory},
  doi={10.1101/2025.04.09.648075},
  url={https://www.biorxiv.org/content/10.1101/2025.04.09.648075v1}
}

@article{Dauparas2022ProteinMPNN,
  title={Robust deep learning--based protein sequence design using ProteinMPNN},
  author={Dauparas, Justas and Anishchenko, Ivan and Bennett, Nathaniel and Bai, Hua and Ragotte, Robert J and Milles, Lukas F and Wicky, Basile IM and Courbet, Alexis and de Haas, Rob J and Bethel, Neville and others},
  journal={Science},
  volume={378},
  number={6615},
  pages={49--56},
  year={2022},
  publisher={American Association for the Advancement of Science}
}

@article{Maaten2008TSNE,
  title={Visualizing data using t-SNE},
  author={Maaten, Laurens van der and Hinton, Geoffrey},
  journal={JMLR},
  volume={9},
  number={Nov},
  pages={2579--2605},
  year={2008}
}

@inproceedings{Lin2023ESMIF,
  title={Learning inverse folding from millions of predicted structures},
  author={Hsu, Chloe and Verkuil, Robert and Liu, Jason and Lin, Zeming and Hie, Brian and Sercu, Tom and Lerer, Adam and Rives, Alexander},
  booktitle={ICML},
  pages={8946--8970},
  year={2022},
  organization={PMLR}
}

@article{Ferruz2022ProtGPT2,
  title={ProtGPT2 is a deep unsupervised language model for protein design},
  author={Ferruz, Noelia and Schmidt, Steffen and H{\"o}cker, Birte},
  journal={Nature Communications}, 
  volume={13},
  number={1},
  pages={4348},
  year={2022},
  publisher={Nature Publishing Group UK London}
}

@inproceedings{Wang2024DPLM,
  title={Diffusion Language Models Are Versatile Protein Learners},
  author={Wang, Xinyou and Zheng, Zaixiang and Ye, Fei and Xue, Dongyu and Huang, Shujian and Gu, Quanquan},
  booktitle={ICML},
  pages={52309--52333},
  year={2024},
  organization={PMLR}
}

@inproceedings{Wang2024DiMA,
  title={Diffusion on Language Model Encodings for Protein Sequence Generation},
  author={Meshchaninov, Viacheslav and Strashnov, Pavel and Shevtsov, Andrey and Nikolaev, Fedor and Ivanisenko, Nikita and Kardymon, Olga and Vetrov, Dmitry},
  year={2025},
  booktitle={ICML}
}

@inproceedings{Satorras2021EGNN,
  title={E (n) equivariant graph neural networks},
  author={Satorras, V{\i}ctor Garcia and Hoogeboom, Emiel and Welling, Max},
  booktitle={ICML},
  pages={9323--9332},
  year={2021},
  organization={PMLR}
}

@article{dauparasAtomicContextconditionedProtein2025a,
  title={Atomic context-conditioned protein sequence design using LigandMPNN},
  author={Dauparas, Justas and Lee, Gyu Rie and Pecoraro, Robert and An, Linna and Anishchenko, Ivan and Glasscock, Cameron and Baker, David},
  journal={Nature Methods},
  pages={1--7},
  year={2025},
  publisher={Nature Publishing Group US New York}
}

@article{chuSparksFunctionNovo2024,
  title={Sparks of function by de novo protein design},
  author={Chu, Alexander E and Lu, Tianyu and Huang, Po-Ssu},
  journal={Nature Biotechnology},
  volume={42},
  number={2},
  pages={203--215},
  year={2024},
  publisher={Nature Publishing Group US New York}
}

@article{kortemmeNovoProteinDesign2024,
  title={De novo protein design—From new structures to programmable functions},
  author={Kortemme, Tanja},
  journal={Cell},
  volume={187},
  number={3},
  pages={526--544},
  year={2024},
  publisher={Elsevier}
}

@article{notinMachineLearningFunctional2024a,
  title={Machine learning for functional protein design},
  author={Notin, Pascal and Rollins, Nathan and Gal, Yarin and Sander, Chris and Marks, Debora},
  journal={Nature Biotechnology},
  volume={42},
  number={2},
  pages={216--228},
  year={2024},
  publisher={Nature Publishing Group US New York}
}

@article{
doi:10.1126/science.ade2574,
  title={Evolutionary-scale prediction of atomic-level protein structure with a language model},
  author={Lin, Zeming and Akin, Halil and Rao, Roshan and Hie, Brian and Zhu, Zhongkai and Lu, Wenting and Smetanin, Nikita and Verkuil, Robert and Kabeli, Ori and Shmueli, Yaniv and others},
  journal={Science},
  volume={379},
  number={6637},
  pages={1123--1130},
  year={2023},
  publisher={American Association for the Advancement of Science}
}

@inproceedings{dauphin2017language,
  title={Language modeling with gated convolutional networks},
  author={Dauphin, Yann N and Fan, Angela and Auli, Michael and Grangier, David},
  booktitle={ICML},
  pages={933--941},
  year={2017},
  organization={PMLR}
}

\end{document}